\documentclass[journal]{IEEEtran}
\usepackage{cite}
\usepackage[cmex10]{amsmath}
\usepackage{graphicx}

\begin{document}

\title{Phase Noise of the Radio Frequency (RF) Beatnote Generated by a Dual-Frequency VECSEL }

\author{Syamsundar~De, Abdelkrim~El~Amili, Ihsan~Fsaifes,~Gr\'egoire~Pillet,~Ghaya~Baili,~Fabienne~Goldfarb,~Mehdi~Alouini,
~Isabelle~Sagnes,~and~Fabien~Bretenaker,~\IEEEmembership{Member,~IEEE}
\thanks{Syamsundar De, Ihsan Fsaifes, Fabienne Goldfarb, and Fabien Bretenaker are with Laboratoire Aim\'e Cotton, CNRS-Universit\'e Paris Sud 11-ENS Cachan, Orsay, France.}
\thanks{Abdelkrim El Amili and Mehdi Alouini are with Institut de Physique de Rennes, CNRS-Universit\'e de Rennes I, Rennes, France.}
\thanks{Gr\'egoire Pillet and Ghaya Baili are with Thales Research and Technology, Palaiseau, France.}
\thanks{Isabelle Sagnes is with Laboratoire de Photonique et Nanostructures, CNRS, Marcoussis, France.}
\thanks{Manuscript received June ??, 2013.}}

\markboth{Journal of lightwave technology,~Vol.~??, No.~??, ??~2013}%
{Shell \MakeLowercase{\textit{et al.}}: Bare Demo of IEEEtran.cls for Journals}

\maketitle

\begin{abstract}
We analyze, both theoretically and experimentally, the phase noise of the radio frequency (RF) beatnote generated by optical mixing of two orthogonally polarized modes in an optically pumped dual-frequency Vertical External Cavity Surface Emitting Laser (VECSEL). The characteristics of the RF phase noise within the frequency range of 10 kHz - 50 MHz are investigated for three different nonlinear coupling strengths between the two lasing modes. In the theoretical model, we consider two different physical mechanisms responsible for  the RF phase noise. In the low frequency domain (typically below 500 kHz), the dominant contribution to the RF phase noise is shown to come from the thermal fluctuations of the semicondutor active medium induced by pump intensity fluctuations. However, in the higher frequency domain (typically above 500~kHz), the main source of RF phase noise is shown to be the pump intensity fluctuations which are transfered to the intensity noises of the two lasing modes and then to the phase noise via the large Henry factor of the semiconductor gain medium. For this latter mechanism, the nonlinear coupling strength between the two lasing modes is shown to play an important role in the value of the RF phase noise. All experimental results are shown to be in good agreement with theory. 
\end{abstract}

\begin{IEEEkeywords}
Dual-frequency laser, VECSELs, phase noise.
\end{IEEEkeywords}

\section{Introduction}

\IEEEPARstart{T}{he} increasing demand for spectrally pure optically-carried radio frequency (RF) signals comes from their broad application areas such as wide-band signal processing \cite{Tonda2006, Rideout2007}, long-range transmission of high purity RF references \cite{Alouini2001, Narbonneau2006}, ultrastable atomic clocks \cite{Knappe2004}, etc. The dual-frequency laser, sustaining oscillation of two orthogonally polarized modes with frequency difference in the RF range, is a promising source for direct generation of such optically-carried high purity RF signals. The advantages of this technique for the direct generation of optically-carried RF signals, compared to other obvious techniques such as optical heterodyning of two independent lasers \cite{Scott1992, Seeds2006}, are the narrow linewidth, continuous frequency tunability, and 100 percent modulation depth of the RF signal. Such tunable dual-frequency operation has been realized for different solid-state lasers \cite{Brunel1997, Alouini1998, Czarny2004}, but the main limitation of these solid-state lasers comes from the relatively strong intensity noise due to relaxation oscillations inherent to their class-B dynamical behavior \cite{Arecchi1984, Taccheo1996}. The dual-frequency VECSEL does not suffer from relaxation oscillations because of its class-A dynamical behavior, which is ensured by achieving longer photon lifetime ($ \sim $10 ns) inside the cm-long external cavity than the carriers' lifetime ($ \sim $3 ns) of the semiconductor gain medium \cite{Baili2009}. Moreover, similarly to dual-frequency solid-state lasers, the RF beatnote generated by the dual-frequency VECSEL is widely tunable by varying the intra-cavity phase anisotropy
 \cite{Baili2009}. 

More importantly, the spectral purity of the beatnote, which depends on the correlation of the intensity and phase fluctuations of the two lasing modes, is very high since the two orthogonal polarization modes are oscillating inside the same cavity. We have already investigated such correlation behavior between the intensity noises of the two modes of dual-frequency VECSEL for different coupling situations \cite{De2013}. Now, we know that the spectral purity of the RF beatnote is actually related to the correlation of the optical phase noises of the two modes, since the positively correlated part of the phase fluctuations will cancel out in the beatnote generated by optical mixing of the two laser modes. But the intensity noise correlation between the two laser modes  is also expected to have strong influence on the spectral purity of the RF beatnote as phase fluctuations are coupled to intensity flcutuations due to the large Henry factor of the semiconductor gain medium of such a VECSEL \cite{Henry1982, Henry1986, Agrawala1989}. Moreover, there are other reasons such as thermal or mechanical fluctuations of the laser cavity, which can create some additional phase noise. 

In the present paper, we analyze, both experimentally and theoretically, the behavior of the phase noise of the RF beatnote for different coupling situations between the two laser modes within the 10 kHz - 50 MHz frequency range. Indeed, the simultaneous oscillation of the two modes in the dual-frequency laser is ruled by the nonlinear coupling constant which must be the lowest possible for robust operation. By contrast, the spectral purity of the beat note is expected to increase if the two modes share exactly the same optical path into the laser, which leads to a high nonlinear coupling constant. Thus, a thorough understanding of how the phase noise of the RF beatnote evolves with respect to the mode coupling strength is useful. In the theoretical model, we suppose that the only source of noise is the intensity noise of the pump diode laser, which is white within the frequency range of our interest (10 kHz - 50 MHz). Moreover, the pump fluctuations entering into the two laser modes are partially correlated, but are in phase. Now in the model we consider two different mechanisms through which the pump intensity noise is affecting the phase noise of the RF beatnote. Firstly, the pump intensity noise is contributing to the higher frequency (typically for frequencies larger than 500 kHz) part of the RF phase noise via phase-intensity coupling due to the large Henry factor of the semiconductor active medium. The other contribution of the pump noise to the RF phase noise (typically for frequencies lower than 500 kHz) is coming through the thermal fluctuations of the refractive index of the semiconductor structure and hence the fluctuations of the effective cavity length \cite{Horak2006}. In Section II, we develop a theoretical model aiming at describing the effects of the pump noise on the phase noise of the RF beatnote. Section III describes the experimental set-up for measuring the RF phase noise. In Section IV, we compare the experimental results with the predictions of our theoretical model for different coupling strengths between the two modes, controlled by the spatial separation of the two modes inside the active medium.  

%

\section{Theoretical Model}
\begin{figure}[]
\centering
\includegraphics[width=2.5in]{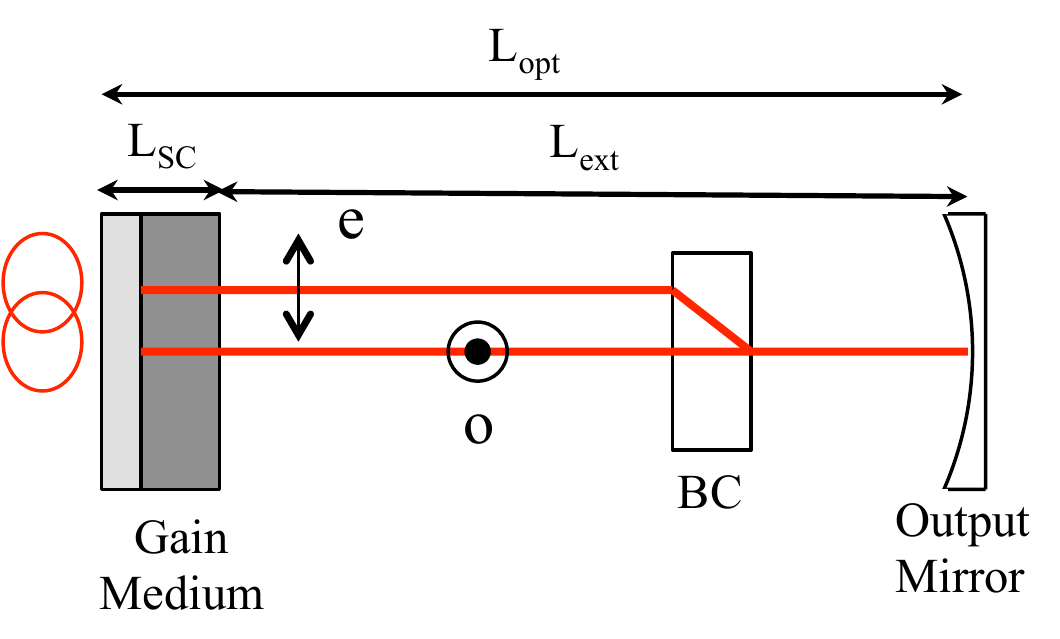}
\caption{Schematic of the basic principle of the dual-frequency VECSEL. The birefringent crystal (BC) spatially separates the two modes polarized along the ordinary (o) and extraordinary (e) directions. They partially overlap in the active medium. The length of the semiconductor 1/2-VCSEL is $L_{SC}$. The length of the external cavity is $L_{ext}$. The optical length of the cavity is $L_{opt}$.}\label{Fig01}
\end{figure}

The basic working principle of our dual-frequency VECSEL, in which two orthogonally polarized modes are partially spatially separated in the active medium by an intra-cavity birefringent element, is schematized in Fig.\ \ref{Fig01}. To model the effect of pump noise on the noise properties of this laser, we take into account two different physical mechanisms. They are described in the following subsections. 
\subsection{Propagation of intensity to phase noise via phase-intensity coupling}
In this subsection, we generalize the model developed earlier \cite{De2013}. This model has proved to be successful to describe the correlations between the intensity noises of the two modes induced by the intensity noise of the pump laser. We now generalize this model to take into account the transfer of intensity noise to the phase noise of the RF beatnote via phase-intensity coupling through the Henry factor. The rate equations governing the laser behavior read \cite{De2013}: 
\begin{align}
&\frac{dF_{1}(t)}{dt}=-\frac{F_{1}(t)}{\tau_{1}}+\kappa N_{1}(t) F_{1}(t) ,\\
&\frac{dF_{2}(t)}{dt}=-\frac{F_{2}(t)}{\tau_{2}}+\kappa N_{2}(t) F_{2}(t) ,\\
&\frac{dN_{1}(t)}{dt}=\dfrac{1}{\tau}[N_{01}(t)-N_{1}(t)]-\kappa N_{1}(t) [F_{1}(t)+\xi_{12}F_{2}(t)] ,\\
&\frac{dN_{2}(t)}{dt}=\dfrac{1}{\tau}[N_{02}(t)-N_{2}(t)]-\kappa N_{2}(t) [F_{2}(t)+\xi_{21}F_{1}(t)] ,\\
&\frac{d\phi(t)}{dt}=\dfrac{d\phi_{1}(t)}{dt}-\dfrac{d\phi_{2}(t)}{dt}=\frac{\alpha}{2} \kappa [N_{1}(t)-N_{2}(t)] .
\end{align}
Here $ F_{1} $ and $ F_{2} $ are the photon numbers in the two cross-polarized modes, $ N_{1} $ and $ N_{2} $ are the corresponding population inversions, $ \phi_{1} $ and $ \phi_{2} $ are the optical phases of the two lasing modes and $ \phi $ is the phase of the RF beatnote, generated by optical mixing of the two laser modes. $ \tau_{1} $, $ \tau_{2} $ are the photon lifetimes inside the cavity for the two modes. We choose to consider two different photon lifetimes, since the two modes experience different losses inside the cavity. $ \tau $ is the population inversion lifetime, which is identical for the two modes. $ N_{01} $ and $ N_{02} $ are the two unsaturated population inversions, since $ { N_{01}}/{\tau} $ and $ { N_{02}}/{\tau} $ denote the pumping rates for the two modes, respectively. The stimulated emission coefficient $ \kappa $ is proportional to the stimulated emission cross-section. $ \alpha $ holds for the Henry factor, which is the ratio of the carrier induced change of real part of refractive index to the corresponding change of imaginary part of refractive index \cite{Henry1982}. The coefficients $\xi_{12} $ and $ \xi_{21} $ are the ratios of the cross- to self-saturation coefficients, which take into account the effect of partial overlap between the two modes inside the gain medium. Therefore, 
\begin{equation}
C=\xi_{12}\xi_{21}
\end{equation}
defines the nonlinear coupling constant \cite{Lamb1989}, on which depends the possibility for the two modes to oscillate simultaneously. This approach will allow us to vary the coupling constant ($ C $) simply by changing the ratios of the cross- to self-saturation coefficients, through adjustment of the spatial overlap between the two modes in the active medium. This technique is very simple, although it is not easy to describe in the framework of the usual spin-flip model \cite{Miguel1995, Travagnin1997, Exter1998, Gahl1999, Kaiser2002}. The steady-state solutions for simultaneous oscillation of the two orthogonally polarized modes, obtained from Eqs. (1-4), are given as 
\begin{align}
&F_{1}\equiv F_{10}=\frac{(r_{1}-1)-\xi_{12}(r_{2}-1)}{\kappa\tau(1-C)} ,\\
&F_{2}\equiv F_{20}=\frac{(r_{2}-1)-\xi_{21}(r_{1}-1)}{\kappa\tau(1-C)} ,\\
&N_{1}\equiv N_{1th}=\frac{1}{\kappa\tau_{1}} ,\\
&N_{2}\equiv N_{2th}=\frac{1}{\kappa\tau_{2}} .
\end{align}
Here $ r_{1}= \overline{N}_{01}/N_{1th} $ and $ r_{2}=\overline{N}_{02}/N_{2th} $ are the excitation ratios for the two modes, where $ \overline{N}_{01} $ and $  \overline{N}_{02} $ are the steady-state values of the corresponding unsaturated population inversions. In the following, we suppose that the pump intensity noise is the dominant source of noise in the considered frequency range (10 kHz - 50 MHz). The pump fluctuations for the two modes are modeled as: 
\begin{align}
N_{01}(t)&=\overline{N}_{01}+\delta N_{01}(t)\ ,\\
N_{02}(t)&=\overline{N}_{02}+\delta N_{02}(t)\ .
\end{align}
The fluctutations in photon numbers and corresponding population inversions around their steady-state values are defined as 
\begin{align}
F_{1}(t)&=F_{10}+\delta F_{1}(t)\ ,\\
F_{2}(t)&=F_{20}+\delta F_{2}(t)\ ,\\
N_{1}(t)&=N_{1th}+\delta N_{1}(t)\ ,\\
N_{2}(t)&=N_{2th}+\delta N_{2}(t)\ .
\end{align}
If now we substitute Eqs. (11-16) into Eqs. (1-4) and then perform Fourier transformation, followed by linearization around the steady-state, we obtain the following expression, relating the fluctuations of the photon numbers of the two laser modes to the pump fluctuations, in the frequency domain:
\begin{equation}
\begin{bmatrix}
\widetilde{\delta F}_{1}(f)\\
\widetilde{\delta F}_{2}(f)
\end{bmatrix}
=
\begin{bmatrix}
M_{11}(f) & M_{12}(f)\\
M_{21}(f) & M_{22}(f)
\end{bmatrix}
\begin{bmatrix}
\tilde{\delta N_{01}}(f)\\
\tilde{\delta N_{02}}(f)
\end{bmatrix}\ ,
\end{equation}
where the tilde symbol indicates Fourier transform and where
\begin{align}
M_{11}(f)&=\frac{1}{\tau}\frac{[\frac{1}{\tau_{2}}-\frac{2i\pi f}{\kappa F_{20}}(r_{2}/\tau - 2i\pi f)]}{\Delta (f)}\ ,\\
M_{12}(f)&=\frac{\xi_{12}}{\tau \tau_{1} \Delta (f)}\ ,\\
M_{21}(f)&=\frac{\xi_{21}}{\tau \tau_{2} \Delta (f)}\ ,\\
M_{22}(f)&=\frac{1}{\tau}\frac{[\frac{1}{\tau_{1}}-\frac{2i\pi f}{\kappa F_{10}}(r_{1}/\tau - 2i\pi f)]}{\Delta (f)}\ ,
\end{align}
and
\begin{align}
\Delta (f)&= [\frac{1}{\tau_{1}}-\frac{2i\pi f}{\kappa F_{10}}(r_{1}/\tau - 2i\pi f)]\nonumber\\
&\times
[\frac{1}{\tau_{2}}-\frac{2i\pi f}{\kappa F_{20}}(r_{2}/\tau - 2i\pi f)]
-C/\tau_{1}\tau_{2}\ .
\end{align}
Therefore, the fluctuations of the RF phase in frequency domain can be easily obtained by Fourier transforming Eq.~(5) and combining it with Eqs. (1) and (2), leading to: 
\begin{equation}
\widetilde{\delta \phi}(f)=\frac{\alpha}{2}[\frac{\widetilde{\delta F}_{1}(f)}{F_{10}}-\frac{\widetilde{\delta F}_{2}(f)}{F_{20}}]\ .
\end{equation}
Now in the experiment, the two modes, which are partially overlapping in the gain medium, are pumped by the same pump laser. Therefore, we choose the following approximations for the pump fluctuations \cite{De2013}:
\begin{equation}
\langle \vert \widetilde{\delta N}_{01}(f) \vert ^{2}\rangle=\langle \vert \widetilde{\delta N}_{02}(f) \vert ^{2}\rangle=\langle \vert \widetilde{\delta N}_{0} \vert ^{2}\rangle\ ,
\end{equation}
\begin{equation}
\langle \widetilde{\delta N}_{01}(f) \tilde{\delta N^{*}_{02}}(f)\rangle=\eta \langle \vert \widetilde{\delta N}_{0} \vert ^{2}\rangle e^{i\psi}\ .
\end{equation}
The pump noises entering into the two laser modes are white noises of identical amplitudes, as implied by Eq. (24). Furthermore, Eq. (25) describes the correlation behavior between the pump fluctuations for the two laser modes. For the theory we have assumed that the pump intensity fluctuations for the two modes are partially correlated ($ 0<\eta< 1$) and $ \eta $  is independent of frequency within the  frequency range of our interest (10 kHz - 50 kHz). Moreover, the pump fluctuations are also assumed to be in phase ($ \psi=0 $). Considering all the above approximations, which have been shown to be valid \cite{De2013}, the power spectral density (PSD) of the phase noise for the RF beatnote can be written as:
\begin{eqnarray}
\langle\vert\widetilde{\delta \phi}(f)\vert^{2}\rangle&=&\frac{\alpha^{2}}{4}[\dfrac{\langle\vert\widetilde{\delta F}_{1}(f)\vert^{2}\rangle}{F^{2}_{10}}+ \dfrac{\langle\vert\widetilde{\delta F}_{2}(f)\vert^{2}\rangle}{F^{2}_{20}}\nonumber \\
&&-\frac{2 \langle Re(\widetilde{\delta F}_{1}(f) \widetilde{\delta F}^{*}_{2}(f)) \rangle}{F_{10}F_{20}}]\ .\label{Eq26}
\end{eqnarray}
So Eq. (26) clearly shows the dependence of the RF phase noise not only on the intensity noises (first 2 terms of right hand side) but also on the correlation between the intensity noises of the two laser modes (last term on the right hand side). 
\subsection{Thermal Fluctuations Induced by the Pump Fluctuations}
Thermal fluctuations can have a deleterous effect on the phase noise of the laser modes. Generally, thermal fluctuations introduce fluctuations of cavity length either by varying the effective refractive index inside the cavity or by the varying the physical length of the cavity. These cavity length fluctuations are translated into fluctuations of the optical phases of the laser modes. In the present section, we start, following Refs. \cite{Tropper2006,Laurain2009,Laurain2010}, by deriving the power spectral density (PSD) of the optical phase noise for a single mode laser induced by such thermal fluctuations. Then we extend this to the case of our dual-frequency VECSEL.  

It is worth mentioning that in this study we will not consider the effect of temperature fluctuations either due to entropy change generated by the spontaneous emission usually in solid-state lasers \cite{Foster2004, Foster2008} or free electron generation-recombination associated with materials defects in the semiconductor structure \cite{Fukuda1993}, which explain the so called $ 1/f $ frequency noise. 
\subsubsection{Single mode laser}
The instantaneous angular frequency of the laser, which is an integer multiple of the free spectral range of the cavity, is given by 
\begin{equation}
\omega (t)=q\frac{\pi c}{L_{opt}(t)}\ ,
\end{equation}
where $ q $ is an integer and $ c $ is the vacuum velocity of the light.  The time varying quantity $L_{opt}(t) $ denotes the total optical length of the laser cavity, which can be expressed as
\begin{equation}
L_{opt}(t)=L_{ext}+n_{SC}L_{SC}\ ,
\end{equation}
where $ L_{ext} $ is the length of the external cavity, and where $ n_{SC} $ and $ L_{SC} $ are the effective refractive index and the thickness of the semiconductor structure. Then the variation of the optical frequency due to total cavity length variation can be given as 
\begin{equation}
\delta \omega_{th}(t)=-\omega (t)\frac{\delta L_{opt}(t)}{L_{opt}(t)} .
\end{equation} 
 The dominant contribution in the thermal fluctuations of the optical length of the cavity comes from the fluctuations of the refractive index of the semiconductor structure \cite{Laurain2009,Laurain2010}, leading to:
 \begin{equation}
 \delta \omega_{th}(t)\simeq -\frac{\omega_{0}}{L_{ext}}\overline{\delta n}(t)L_{SC},
\end{equation}  
where $ \overline{\delta n}(t) $ is the temporal fluctuations of the refractive index, induced thermally and averaged over the volume of the optical mode. 
We can describe it as 
\begin{equation}
\delta \omega_{th}(t)\simeq -\omega_0\Gamma_{th}\overline{\delta T}(t) ,
\end{equation}
where $ \overline{\delta T}(t) $ denotes the fluctuation of the temperature, averaged over the optical mode volume and $ \Gamma_{th} $ is defined as 
\begin{equation}
\Gamma_{th}=\dfrac{L_{SC}}{L_{ext}}\frac{d \overline {n}}{d T}.
\end{equation}
In general, the temperature fluctuation $ \overline{\delta T}(t) $ of the semiconductor structure averaged over the optical mode volume is the sum of two terms \cite{LaurainThese}:
\begin{equation}
  \overline{\delta T}(t)= {\delta T_{k}}(t)+\overline{R_{th}}\delta P_{p}(t)*\theta (t) , \label{Eq33}
\end{equation}
where $ {\delta T_{k}}(t) $ defines the  fundamental thermodynamic fluctuations of the semiconductor, when it is maintained at a fixed temperature $ T $ and the last term of the right hand side describes the effect of temperature fluctuations induced by pump power fluctuations.$ * $' represents the convolution product. $ \overline{R_{th}} $ is the thermal impedance of the semiconductor structure, averaged over the volume of the optical mode. $ \delta P_{p} $ holds for the fluctuations of the pump power and $ \theta (t) $ defines the impulse temperature response of the semiconductor structure to the pump power. As usually performed \cite{Laurain2010}, we model the transfer function $ \Theta (f) $, which is Fourier transform of $ \theta (t) $, as a first-order low-pass filter, leading to:
\begin{equation}
\vert\Theta (f)\vert^{2}=\frac{1}{1+(2\pi f \tau_{th})^{2}} .\label{Eq34}
\end{equation}
The order of magnitude of the response time ($ \tau_{th} $) of thermal diffusion is given as \cite{Davis1998}  
\begin{equation}
\tau_{th}\simeq\frac{w_{p}^{2}}{2\pi D_{T}} , \label{Eq35}
\end{equation}
where $ w_{p} $ is the waist of the pump beam on the semiconductor medium and $D_{T}$ is the thermal diffusion coefficient.
To evaluate the order of magnitude of the first term in the right-hand side of Eq.\; (\ref{Eq33}), we obtain the variance of the intrinsic temperature fluctuations of the semiconductor due to the fundamental thermodynamic fluctuations from Refs. \cite{Chui1992, Gorodetsky2004, Lauer2005}:
\begin{equation}
\langle \delta T_{k}^{2} \rangle=\frac{k_{B}T^{2}}{C_{v}V} .
\end{equation}
Here $ k_{B} $ is Boltzman's constant, $ C_{v} $ is density of specific heat and $ V \simeq \pi w_{0}^{2}L_{\mu c} $ represents the volume of the semiconductor, occupied by the optical mode ($ w_{0} $ is the laser mode size). Therefore, the power spectral density (PSD) of the thermal fluctuations, induced by the fundamental thermodynamic fluctuations, is given as following 
\begin{equation}
S_{T}(f)=\frac{k_{B}\tau_{th}T^{2}}{\pi^{2}f^{2}C_{v}V}\vert\Theta(f)\vert^{2} .
\end{equation}
Similarly the power spectral density of the thermal fluctuations, induced by the pump power fluctuations, is given by 
\begin{equation}
S_{P}(f)= \frac{\overline{R_{th}}^{2} }{4\pi^{2}f^{2}}\vert\Theta(f)\vert^{2}\mathrm{RIN}_{P}(f)P^{2}_{P} ,
\end{equation}
where $ \mathrm{RIN}_{P} $ is the pump relative intensity noise (RIN) and $ P_{P} $ is the incident pump power. For our laser, if we compare the PSD of the thermodynamic fluctuations ($ S_{T}(f) $) with the PSD of the thermal fluctuations, induced by the pump power fluctuations ($ S_{P}(f) $), we find that the effect of the fundamental thermodynamic fluctuations is very small ($ \sim 10^{5} $ times smaller) compared with the effect of the pump power fluctuations. Therefore, in the following, we neglect the effect of fundamental thermodynamical fluctuations, i. e., the first term in the right-hand side of Eq.\ (\ref{Eq33}). 
We can then Fourier transform Eq.\ (\ref{Eq33}) and obtain the optical phase noise power spectral density as
 \begin{eqnarray}
|\widetilde{\delta\phi}_{opt}(f)|^2&=&\dfrac{\omega_{0}^{2} \Gamma^{2}_{th}\overline{R_{th}}^{2}}{4\pi^{2}f^{2}}\vert\Theta(f)\vert^{2}\mathrm{RIN}_{P}(f)P^{2}_{P}\nonumber\\
&=&\vert\Lambda(f)\vert^{2} \vert \widetilde{\delta P}_{P}(f) \vert^{2} ,
\end{eqnarray}
where we have defined the transfer function
\begin{equation}
 \vert\Lambda(f)\vert^{2}=\dfrac{\omega_{0}^{2} \Gamma^{2}_{th}\overline{R_{th}}^{2}}{4\pi^{2}f^{2}}\vert\Theta(f)\vert^{2} ,
 \end{equation}
  and where
 \begin{equation}
 \vert \widetilde{\delta P}_{P}(f) \vert^{2}=\mathrm{RIN}_{P}(f)P^{2}_{P} 
 \end{equation}
  holds for the power spectral density of the pump power fluctuations. 
  
\begin{figure*}[]
\centering
\includegraphics[width=0.7\textwidth]{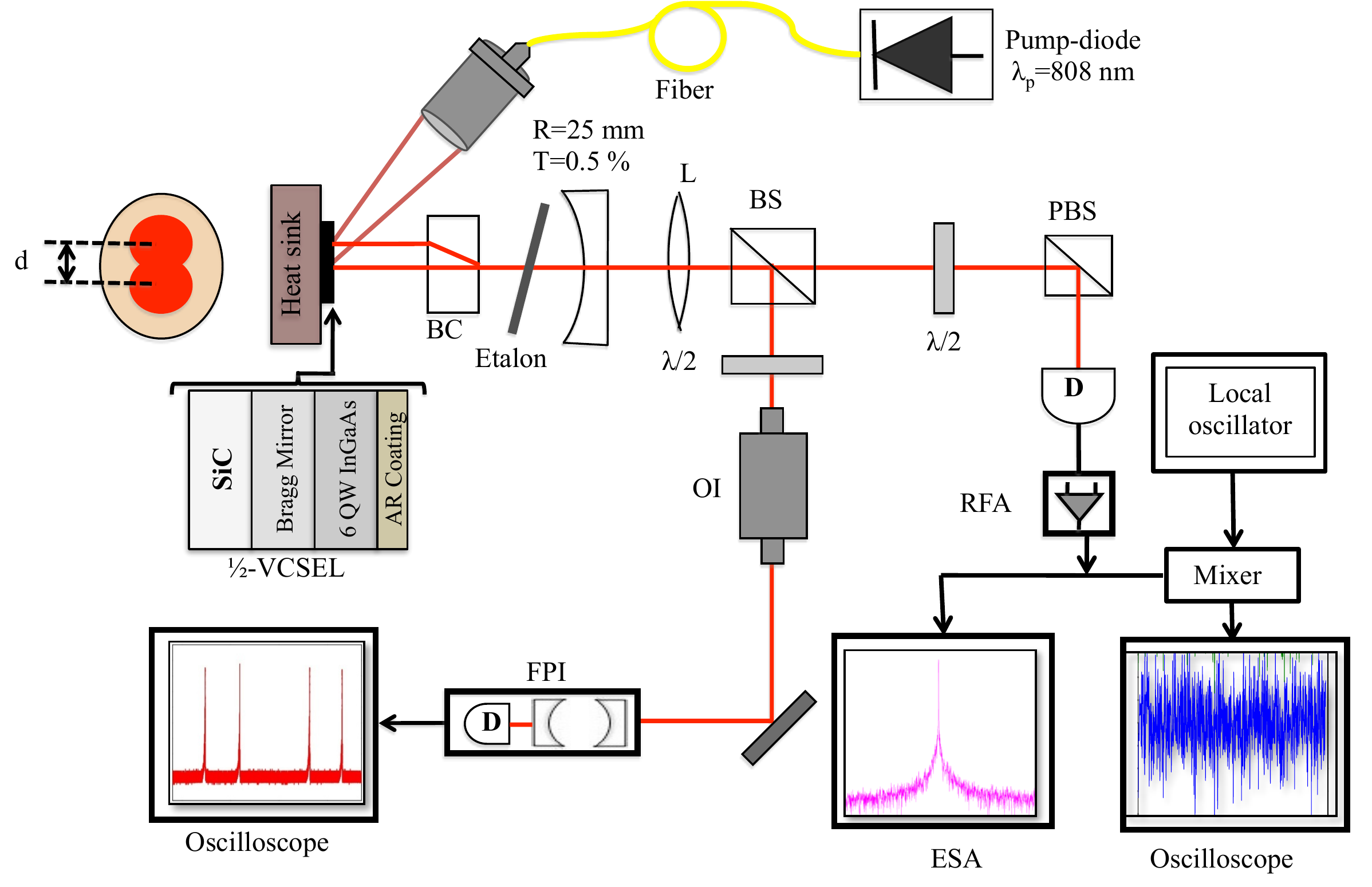}
\caption{Schematic of the experimental set-up for measuring the phase noise of the RF beatnote. $d$: spatial separation between the two modes on the gain structure, BC: birefringent crystal, L: lens, $T$: transmittance of the output coupler, $R$: radius of curvature of the output mirror, BS: beam splitter, $ \lambda/2 $: half-wave plate, PBS: polarization beam spliter, D: photodetector, RFA: radio frequency amplifier, ESA: electrical spectrum analyzer, OI: optical isolator, FPI: Fabry-Perot interferometer.}\label{Fig02}
\end{figure*}
  
\subsubsection{Dual-frequency VECSEL}
In the case of our dual-frequency VECSEL, the two orthogonally polarized modes, having a frequency difference in the RF range, are oscillating inside the same cavity. Therefore, in the frequency domain, the phase fluctuations of the RF beatnote generated by optical mixing of the two laser modes, are given by 
\begin{eqnarray}
\widetilde{\delta \phi}_{RF}(f)&=&\widetilde{\delta \phi}_{1}(f)-\widetilde{\delta \phi}_{2}(f)\nonumber\\
&=&\Lambda_{1}(f)\widetilde{\delta P}_{P1}(f)-\Lambda_{2}(f)\widetilde{\delta P}_{P2}(f) ,
\end{eqnarray}
where $ \widetilde{\delta \phi}_{1}(f) $, $ \widetilde{\delta \phi}_{2}(f) $ are fluctuations of the optical phases of the two laser modes and $ \widetilde{\delta \phi}_{RF}(f) $ is the corresponding RF phase fluctuation in the frequency domain. Moreover, $\widetilde{\delta P}_{P1}(f)$ and $ \widetilde{\delta P}_{P2}(f) $ are the pump fluctuations for the two corresponding laser modes. Now, we make similar approximations for the fluctuations of the pumping rates as in  Eqs. (24-25), leading to: 
\begin{equation}
\vert \widetilde{\delta P}_{P1}(f) \vert^{2}=\vert \widetilde{\delta P}_{P2}(f) \vert^{2}=\vert \widetilde{\delta P}_{P} \vert^{2} ,
\end{equation}
\begin{eqnarray}
\langle \vert \widetilde{\delta P}_{P1}(f) \widetilde{\delta P}^{*}_{P2}(f) \vert \rangle=\eta \vert \widetilde{\delta P}_{P} \vert^{2}\ .
\end{eqnarray}
Eqs. (43) and (44) describe white pumping noises of identical amplitudes for the two laser modes and which are partially uncorrelated ($ 0< \eta < 1 $, $ \eta $ independent of frequency), but in phase ($ \psi =0 $). Now we suppose that 
\begin{equation}
\vert\Lambda_{1}(f)\vert^{2}=\vert\Lambda_{2}(f)\vert^{2}=\vert\Lambda(f)\vert^{2} .
\end{equation}
Therefore, the power spectral density (PSD) of the phase noise of the RF beatnote in our dual-frequency VECSEL due to thermal noise, induced by pump power fluctuations, is found to be
\begin{equation}
\vert\widetilde{\delta\phi}_{RF}(f)\vert^{2}=2(1-\eta)\vert\Lambda(f)\vert^{2}\vert \widetilde{\delta P}_{P} \vert^{2} .\label{Eq46}
\end{equation}
Thus we can conclude that the phase noise of the RF beatnote due to the thermal fluctuations of the refractive index of the semiconductor medium, induced by the pump intensity fluctuations, does not only depend on the PSD of the pump noise, but also on the correlation ($ \eta $) between the pump fluctuations entering into the two laser modes. More specifically, the pump noise induced thermal fluctuations  provide a significant contribution to the RF phase noise as long as the pump fluctuations for the two laser modes are not perfectly correlated ($ \eta <1$).

\section{Description of the experimental set-up}

The experimental set-up designed for measuring the phase noise of the RF beatnote, generated by optical mixing of the two orthogonally polarized laser modes, is schematized in Fig.\ \ref{Fig02}. The laser is operating at 1 $ \mu $m and the pumping is done with a 808 nm diode laser, pigtailed with a multimode fiber and delevering up to 3 W of optical power. The laser is based on the 1/2-VCSEL structure, which is grown by Metal Organic Chemical Vapour Deposition (MOCVD) technique \cite{Laurain2009}. The 1/2-VCSEL structure is initially grown on a GaAs substrate, then it is transferred to a SiC substrate. The high thermal conductivity (490 $ Wm^{-1}K^{-1} $) of the SiC substrate enables to dissipate the extra heat generated by the optical pump and thus maintains a good efficiency of our laser. Moreover, the structure is placed on a Peltier thermoelectric cooler, whose temperature is maintained at $ 19^{\circ} $C using a temperature controller. There is a Bragg mirror inside the 1/2-VCSEL structure,  which consists in 28 pairs of alternating GaAs/AlAs layers, providing 99.9~\% reflectivity at 1 $ \mu $m. The 1/2-VCSEL structure contains six strained balanced InGaAs/GaAsP quantum wells (QW), which are responsible for the gain at 1 $ \mu $m. The structure possesses a linear gain dichroism. The anti-reflection coating of Si$ _{3} $N$ _{4} $ on top of the 1/2-VCSEL structure reduces the micro-cavity effect between the Bragg mirror and the semiconductor-air interface. The planar-concave laser cavity is about 1.5 cm long. A concave mirror of reflectivity 99.5 percent and radius of curvature 25 mm serves as the output coupler. Besides, this configuration ensures a class-A dynamical behavior of our VECSEL, since the photon lifetime ($ \sim $10 ns) is sufficiently longer than the population inversion lifetime ($ \sim $3 ns)\cite{Baili2009, De2013}. The intra-cavity YVO$ _{4} $ birefringent crystal (BC) with anti-reflection coating at 1 $ \mu $m introduces a polarization walk-off $ d $  proportional to its thickness. The two laser modes are perfectly spatially overlapped between the birefringent crystal BC and the output coupler as the Bragg mirror is planar and the output mirror is concave. The birefringent crystal is oriented in such a way that the maximum gain axis of the semiconductor structure is aligned along the bisector of the ordinary and extraordinary eigenpolarizations of the BC. Thanks to the intra-cavity BC, which spatially separates the two orthogonally linearly polarized laser modes on the gain structure while enabling a balanced gain for the two orthogonal polarizations, simultaneous oscillation of the two modes is obtained by reducing the coupling constant value below unity \cite{Baili2009}. 

The waists of the two laser modes are identical and about 62 $ \mu $m. The position and the size of the pump beam is experimentally adjusted on the gain medium to provide maximum and nearly identical gain for the two modes. We have used three different BCs of thicknesses 1 mm, 0.5 mm and 0.2 mm, which correspond to spatial separations  of 100 $ \mu $m, 50 $ \mu $m and 20 $ \mu $m respectively, hence different coupling strengths between the laser modes \cite{Pal2010}. The intra-cavity 150-$ \mu $m-thick uncoated glass \'etalon forces both polarizations to oscillate in single longitudinal mode. 

The scanning Fabry-Perot interferometer with 10 GHz free spectral range enables us to analyze continuously the laser spectrum to check that the two orthogonal eigenpolarizations are oscillating simultaneously in single-longitudinal-mode regime without any mode hopping during data acquisition. To generate the RF beatnote, the two orthogonally polarized modes are mixed by a half-wave plate ($ \lambda/2 $-plate), followed by a polarization beam splitter (PBS). The RF beatnote is detected by a high-speed photodiode of 22 GHz bandwidth (Discovery DSC-30S). Then the RF signal is amplified using a low noise RF amplifier (RFA). Now one part of the amplified RF signal is sent to an electrical spectrum analyzer (ESA) to monitor the beatnote spectrum. Thereafter, the other part of the RF signal is downshifted to an intermediate frequency (IF) by mixing it with a signal from a local oscillator (LO) (Rohde \& Schwarz SMF100A) having very low phase noise compared to the RF phase noise of our laser. Finally the IF signal is recorded in time domain using a deep memory digital oscilloscope and processed numerically to obtain the phase noise spectrum of the RF beatnote. 

\begin{figure}[]
\centering
\includegraphics[width=0.5 \textwidth]{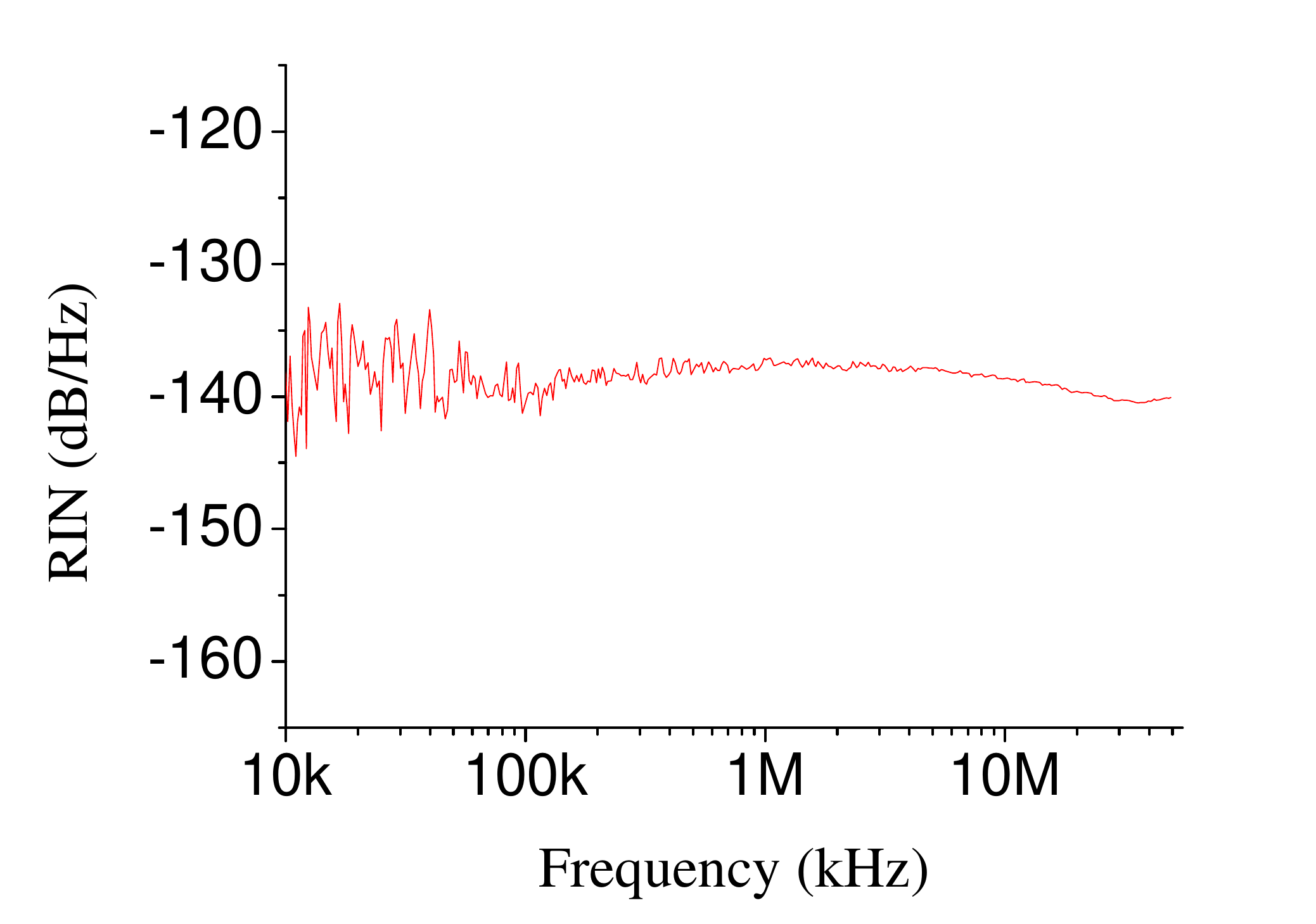}
\caption{Experimentally measured RIN spectrum of the pump diode laser.}\label{Fig06}
\end{figure}

Before measuring the phase noise of the RF beatnote generated by our dual-frequency VECSEL, let us first recall that, in our model, we have approximated the pump intensity noise as white, i.e., independent of frequency in the range of our interest (10 kHz to 50 MHz). Let us first check this assumption by measuring the pump RIN spectrum. The corresponding experimental result is shown in Fig.\ \ref{Fig06}. One can see that the experiment ensures that the pump noise is almost flat within the interesting frequency range (10 kHz $ - $ 50 MHz). The pump RIN level is close to -138 dB/Hz, which we will take as the value that we will inject in our model in the following. 

\section{ RF phase noise: Experimental and theoretical}
The phase noise measurements of the RF beatnote have been performed for three different BCs of thicknesses 1 mm, 0.5 mm and 0.2 mm, which respectively correspond to spatial separations of 100 $ \mu $m, 50 $ \mu $m, and 20 $ \mu $m between the two laser modes on the gain structure. For all the measurements of different coupling situations between the laser modes, the pump power is fixed to 1.5 W and the power of the each laser mode is between 50 and 60~mW.
\begin{figure*}[]
\centering
\includegraphics[width=0.4 \textwidth]{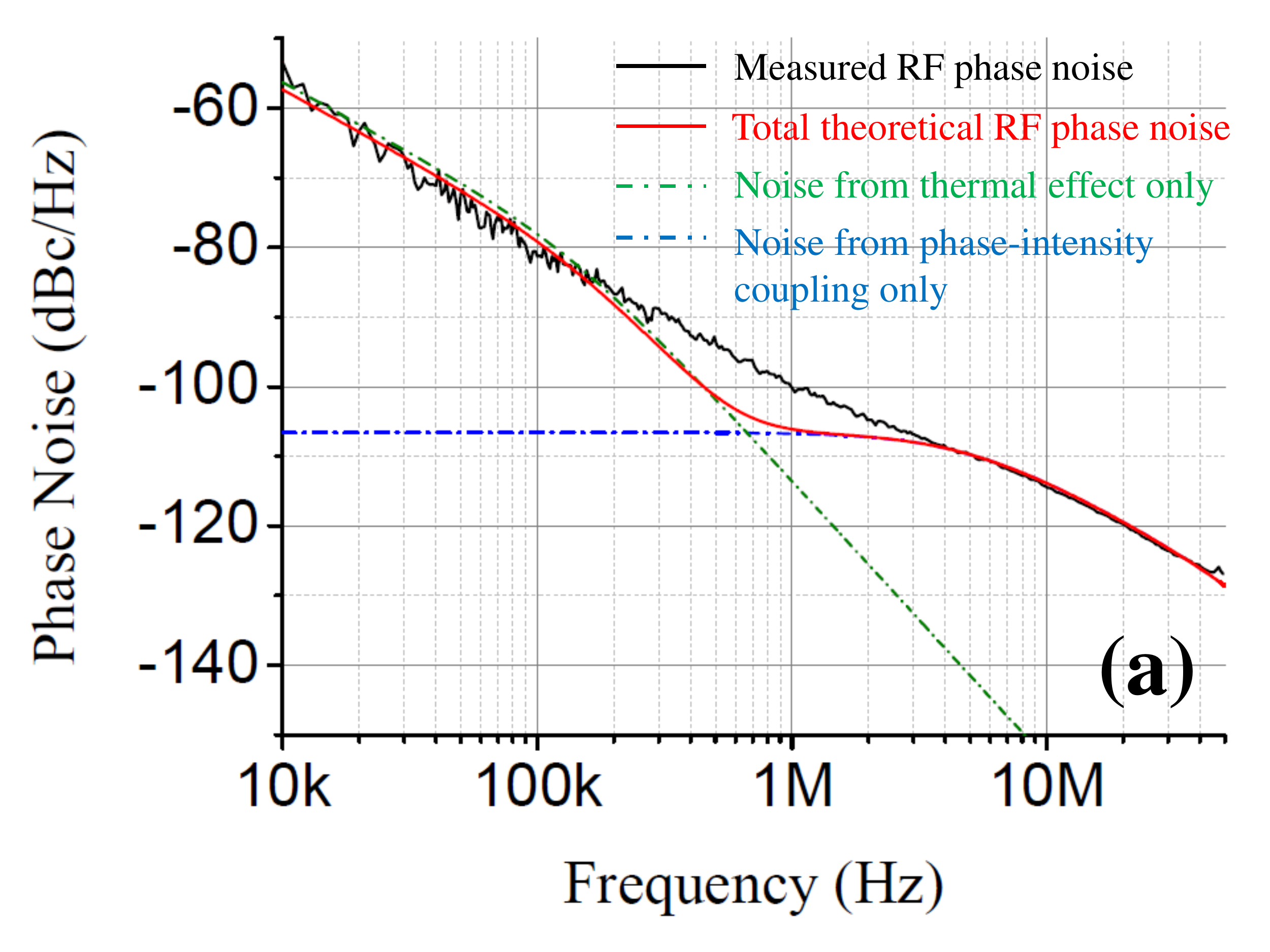}
\includegraphics[width=0.4 \textwidth]{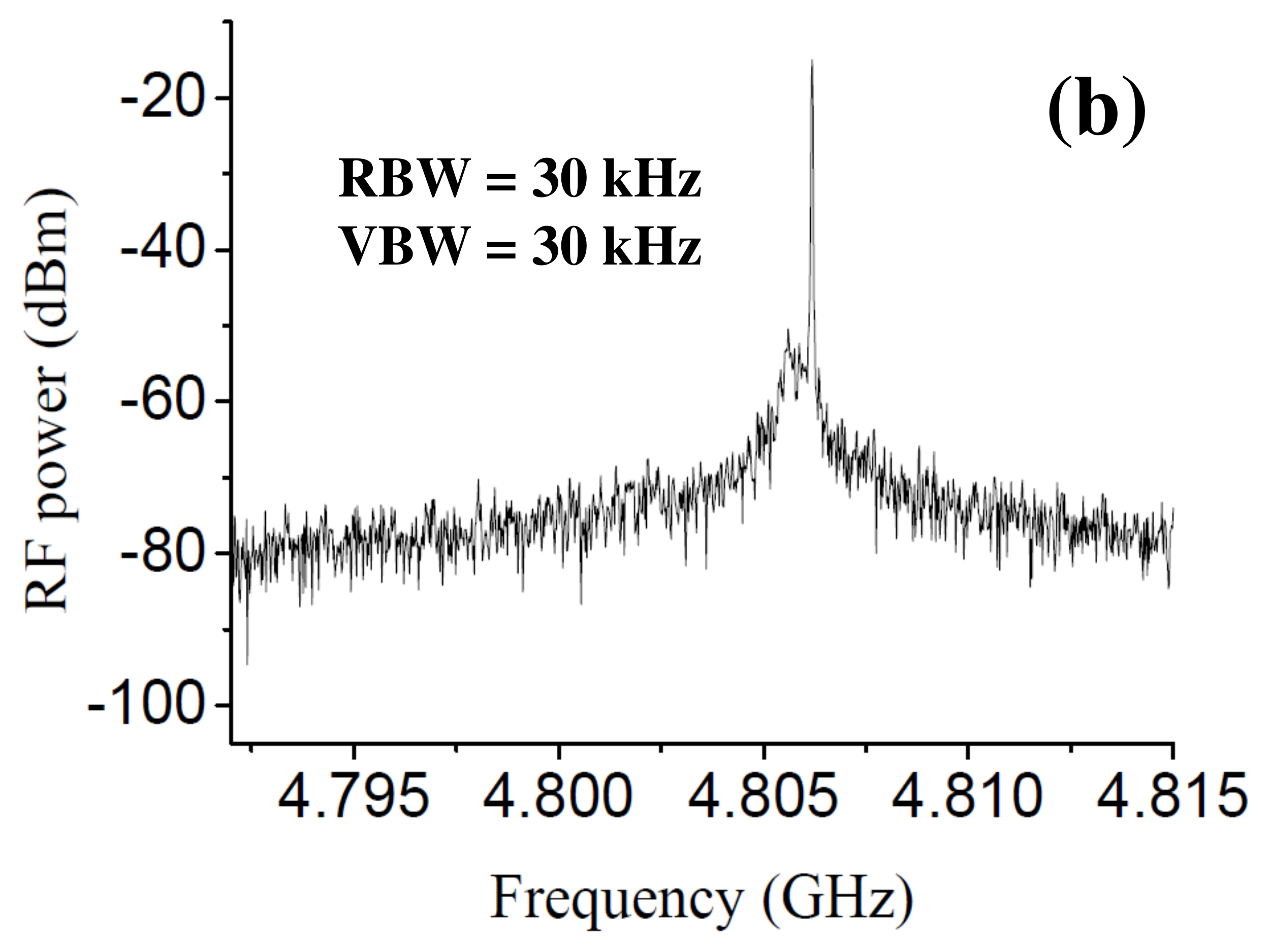}
\caption{Results for the 1-mm-thick BC, which corresponds to a spatial separation ($ d $) of 100 $ \mu $m. (a) Power spectral density (PSD) of the phase noise of the RF beatnote. (b) Beatnote spectrum, measured by the ESA with  resolution bandwidth (RBW) = 30 kHz and video bandwidth (VBW) = 30 kHz. Parameter values used for simulation: $ C=0.1 $, $ \alpha=12 $, $r _{1} = 1.45$, $ r_{2}=1.5 $, $ \tau_{1}= 5.8 $ ns, $ \tau_{2}= 6 $ ns, $ \tau = 3 $ ns, $ RIN_{p}=-138 $ dB/Hz, $ \eta = 0.85 $,  $ D_{T}=0 .22 $ cm$ ^{2} $s$^{-1} $, $ d\overline{n}/{dt}= 2.7\times 10^{-4}$ K$ ^{-1} $, $ L_{SC} = 2.3$ $ \mu $m, $ L_{ext}=1.5 $ cm, $ w_{p}=50 $ $ \mu $m, $ R_{th}=6 $ K.W$ ^{-1} $, $ P_{p}=1.5 $ W. }\label{Fig03}
\end{figure*}

Fig.\ \ref{Fig03} shows the results for the 1-mm thick BC, which spatially separates the two laser beams by 100 $ \mu $m on the gain medium. In this case, the coupling constant ($ C $) has been measured to be equal to 0.1 \cite{Pal2010, De2013}. The  power spectral density (PSD) of the RF phase noise is shown in Fig.\ \ref{Fig03}(a), whereas the corresponding RF beatnote spectrum is reproduced in Fig.\ \ref{Fig03}(b). In Fig.\ \ref{Fig03}(a), the solid black  and red curves respectively correspond to the experimental and total theoretical PSD of the RF phase noise. The total theoretical RF phase noise PSD [red curve in Fig.\ \ref{Fig03}(a)] is obtained by adding two different physical mechanisms: (i) phase-intensity coupling due to the large Henry factor of the semiconductor gain medium of our laser [see Eq. (\ref{Eq26})] and (ii) the fluctuations of the refractive index of the semiconductor structure due to the temperature fluctuations induced by pump intensity fluctuations [see Eq. (\ref{Eq46})]. The contribution of the phase-intensity coupling to the RF phase noise PSD, calculated from Eq. (\ref{Eq26}), is shown by the dot-dashed blue curve in Fig.\ \ref{Fig03}(a), while the contribution of the thermal effect, calculated from Eq. (\ref{Eq46}), is plotted  as the dot-dashed green curve. The parameters used for the simulations, which are given in the figure captions, have been either obtained from the experiment or from preceding works performed with the same type of structure \cite{LaurainThese}.
So in this low coupling condition ($ C=0.1 $), the dominant contribution to the RF phase noise for offset frequencies between 700 kHz and 50 MHz comes from phase-intensity coupling. The RF phase noise coming from thermal fluctuations, induced by pump power fluctuations, is dominant at lower frequencies (between 10 and 700 kHz) but is filtered out for frequencies larger than a few hundred kilohertz by the thermal  response time of the structure.

The total theoretical curve (solid red curve in Fig.\ \ref{Fig03}(a)) shows satisfactory matching with the experimentally measured RF phase noise (solid black curve in Fig.\ \ref{Fig03}(a)). The remaining small discrepancy probably comes from the crude approximation of the thermal response of the structure by a simple first-order filter [see Eq. (\ref{Eq34})] and also from the approximate expression [see Eq. (\ref{Eq35})] we chose for the thermal response time of the structure \cite{Davis1998}. In the beatnote spectrum as shown in Fig.\ \ref{Fig03}(b), one can clearly see that the beat frequency, centered around 4.807 GHz, is sitting on a noise pedestal corresponding to the phase noise of Fig.\ \ref{Fig03}(a).

\begin{figure*}[]
\centering
\includegraphics[width=0.4 \textwidth]{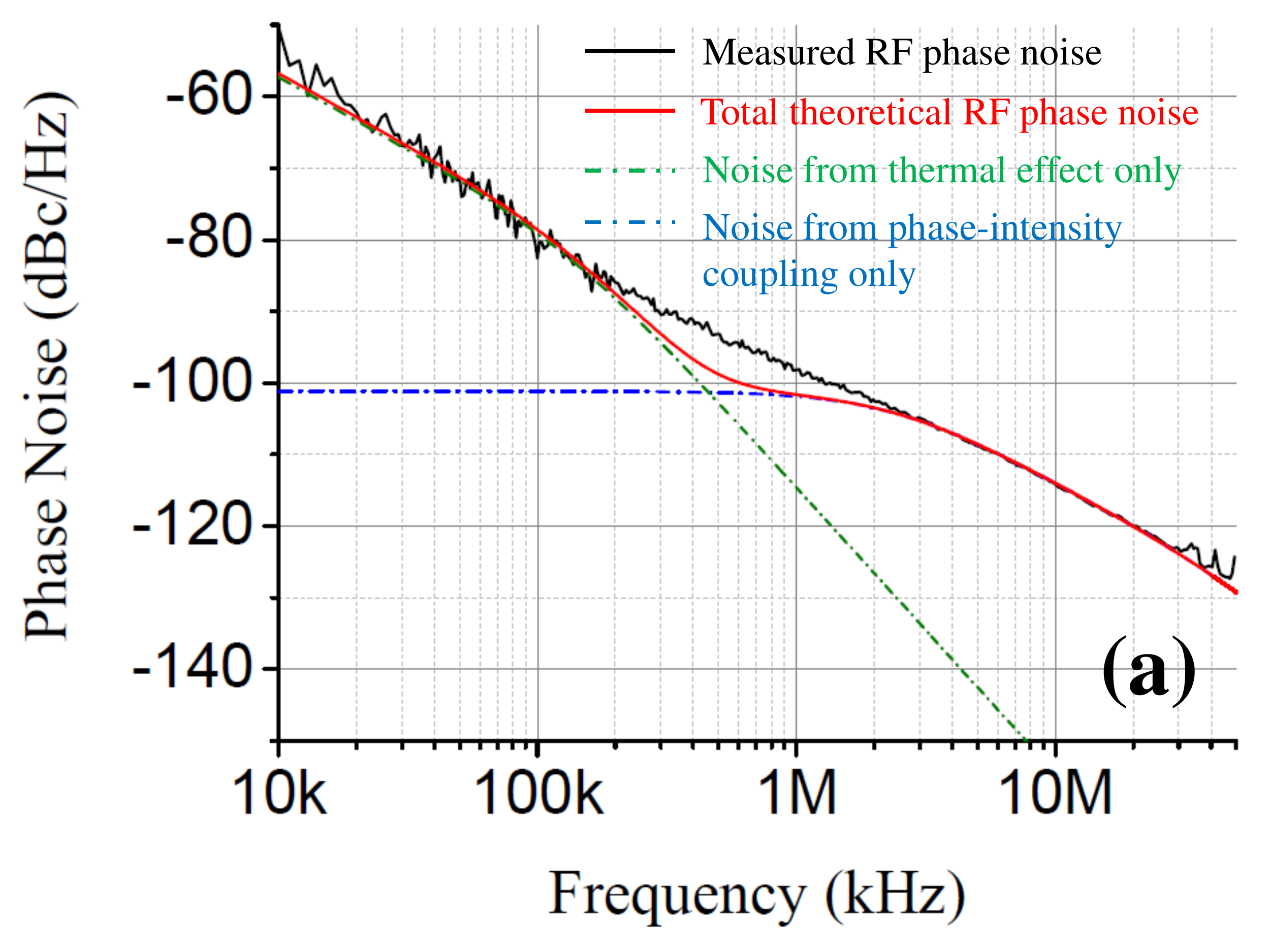}
\includegraphics[width=0.4 \textwidth]{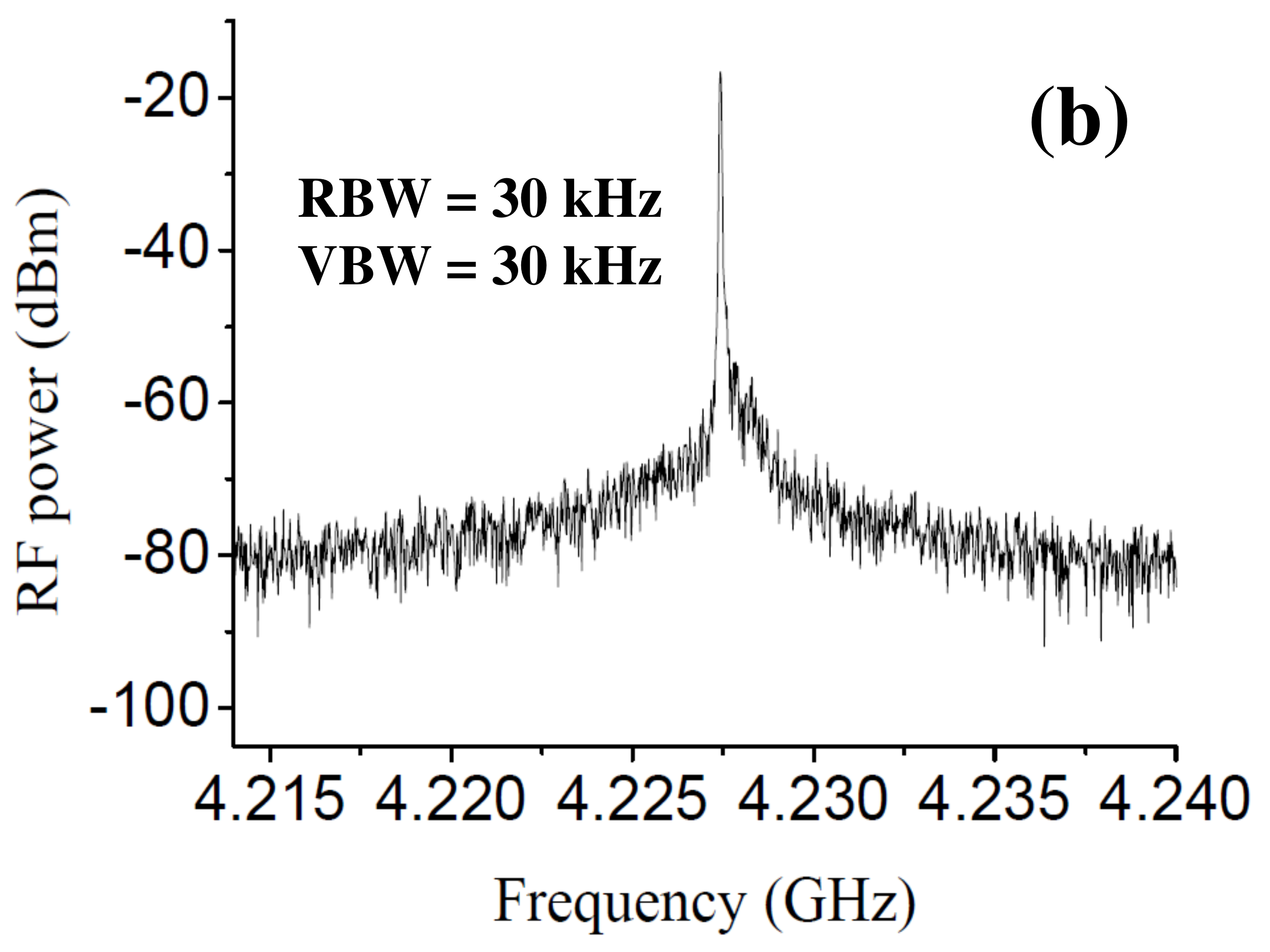}
\caption{Results for the 0.5-mm-thick BC, which corresponds to a spatial separation ($ d $) of 50 $ \mu $m. (a) Power spectral density (PSD) of the phase noise of the RF beatnote. (b) Beatnote spectrum, measured by ESA with resolution bandwidth (RBW) = 30 kHz and video bandwidth (VBW) = 30 kHz. Parameter values used for simulation: $ C=0.35 $, $ \alpha=12 $, $r _{1} = 1.45$, $ r_{2}=1.5 $, $ \tau_{1}= 5.8 $ ns, $ \tau_{2}= 6 $ ns, $ \tau = 3 $ ns, $ RIN_{p}=-138 $ dB/Hz, $ \eta = 0.85 $,  $ D_{T}=0 .22 $ cm$ ^{2} $s$^{-1} $, $ d\overline{n}/{dt}= 2.7\times 10^{-4}$ K$ ^{-1} $, $ L_{SC} = 2.3$ $ \mu $m, $ L_{ext}=1.5 $ cm, $ w_{p}=50 $ $ \mu $m, $ R_{th}=6 $ K.W$ ^{-1} $, $ P_{p}=1.5 $ W. }\label{Fig04}
\end{figure*}

Fig.\ \ref{Fig04} shows the results obtained with the 0.5-mm-thick BC, which corresponds to a spatial separation of 50 $ \mu $m between the two laser modes on the gain structure. For the theoretical simulations, we use the coupling constant ($ C $) value of 0.35, to which we refer as a moderate coupling situation \cite{Pal2010, De2013}. In Fig.\ \ref{Fig04}(a), the solid black curve represents the experimental PSD of the RF phase noise, whereas the solid red curve depicts the corresponding theoretical one. The total theoretical curve (solid red curve) is obtained by considering the effects of thermal noise as well as noise due to phase-intensity coupling. The PSD of the RF phase noise coming  only from pump noise induced thermal noise, is shown as a dot-dashed green curve of Fig.\ \ref{Fig04}(a), whereas the contribution of the phase-intensity coupling due to the Henry factor is given by the dot-dashed blue curve of Fig.\ \ref{Fig04}(a). As we can see from Fig.\ \ref{Fig04}(a), the effect of the phase-intensity coupling is dominant for frequencies higher than 500 kHz, whereas the thermal noise is dominant for lower frequencies  (10 kHz to 500 kHz).  The theoretical prediction (red curve of Fig.\ \ref{Fig04}(a)) shows fairly good agreement with the corresponding experimental results (black curve of Fig.\ \ref{Fig04}(a)). The RF beatnote centered at about 4.227 GHz is shown in Fig.\ \ref{Fig04}(b). 
  
\begin{figure*}[]
\centering
\includegraphics[width=0.4 \textwidth]{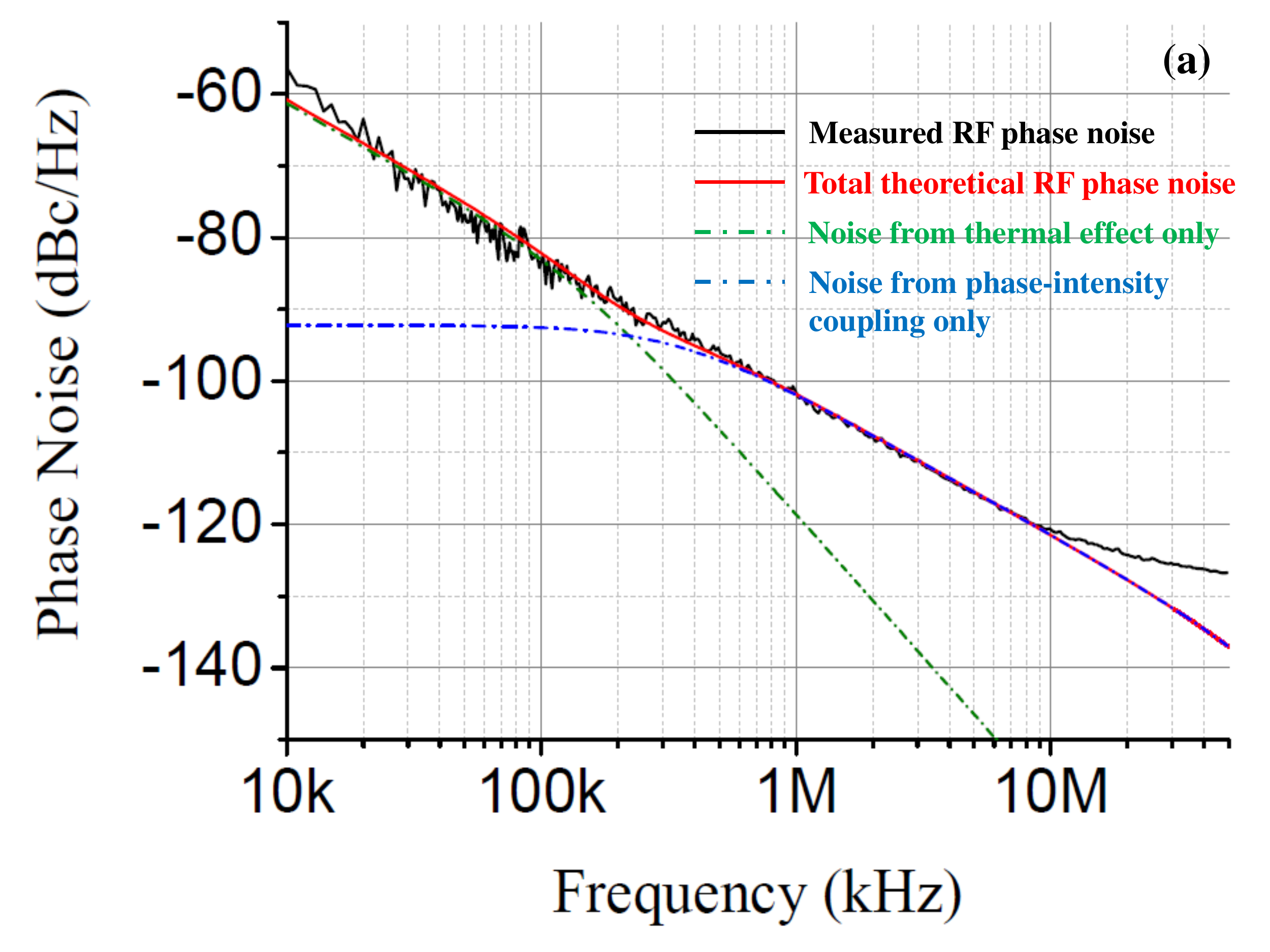}
\includegraphics[width=0.4 \textwidth]{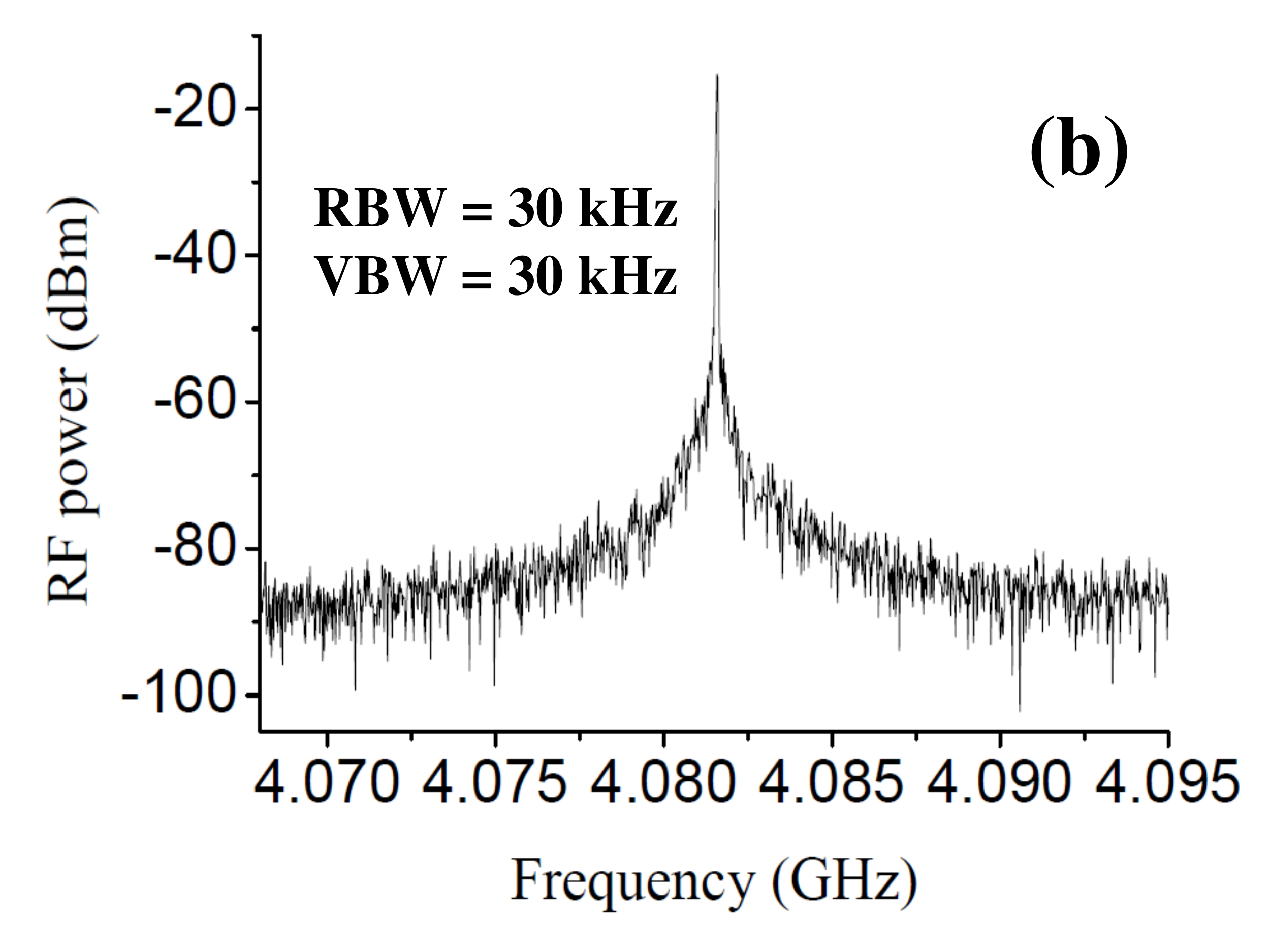}
\caption{Results for the 0.2-mm-thick BC, which corresponds to a spatial separation ($ d $) of 20 $ \mu $m. (a) Power spectral density (PSD) of the phase noise of the RF beatnote. (b) Beatnote spectrum, measured by ESA with resolution bandwidth (RBW) = 30 kHz and video bandwidth (VBW) = 30 kHz. Parameter values used for simulation: $ C=0.65 $, $ \alpha=12 $, $r _{1} = 1.35$, $ r_{2}=1.4 $, $ \tau_{1}= 6.3 $ ns, $ \tau_{2}= 6.6 $ ns, $ \tau = 3 $ ns, $ RIN_{p}=-138 $ dB/Hz, $ \eta = 0.85 $,  $ D_{T}=0 .22 $ cm$ ^{2} $s$^{-1} $, $ d\overline{n}/{dt}= 2.7\times 10^{-4}$ K$ ^{-1} $, $ L_{SC} = 2.3$ $ \mu $m, $ L_{ext}=1.5 $ cm, $ w_{p}=50 $ $ \mu $m, $ R_{th}=6 $ K.W$ ^{-1} $, $ P_{p}=1.5 $ W. }\label{Fig05}
\end{figure*}

The results for the 0.2-mm-thick BC are shown in Fig.\ \ref{Fig05}, which corresponds to a spatial separation between the two laser modes of 20 $ \mu $m. In this configuration, the coupling constant $C$ is shown to be equal to 0.65 \cite{De2013, Pal2010}. So this leads to a relatively intense coupling between the laser modes. In this case, the experimental and theoretical PSD of the RF phase noise within the interesting frequency range 10 kHz $ - $ 50 MHz, are respectively given by the solid black and red curves in Fig.\ \ref{Fig05}(a). Similarly to preceding results, the total theoretical curve (solid red curve of Fig.\ \ref{Fig05}(a)) is obtained by taking into account the two different mechanisms discussed above. We can see that the lower frequency (10 kHz $ - $ 200kHz) RF phase noise is mainly coming from pump noise induced thermal fluctuations, whereas the phase-intensity coupling effect is contributing dominantly for frequencies higher than 200 kHz. The beatnote spectrum is given in Fig.\ \ref{Fig05}(b), where we can see that the beatnote is centered around 4.081 GHz and contains again a few megahertz wide pedestal due to the phase noise. Again the theory (solid red curve) shows very good agreement with experiment (solid black curve) for this strong coupling situation. The little discrepancy for frequencies higher than 20 MHz is coming from the fact that we are approaching the measurement noise floor ($ \sim  - 130\;\mathrm{dBc/Hz}$), which is limited by our oscilloscope. 

We can see from the result obtained in Sec. II that the thermally induced phase noise is independent of the value of the coupling constant $C$, whereas the phase noise coming through phase-intensity coupling via Henry factor depends on $C$ since the intensity noises of the two modes and their correlations depend on $ C $ \cite{De2013}. So for very strong coupling ($ C=0.65 $), the intensity noises of the two laser modes are increased for intermediate frequencies  (between 200~kHz and 2~MHz) due to strong anti-phase relaxation mechanism compared to in-phase relaxation mechanism of the coupled oscillator system \cite{De2013}. As a result of that the RF phase noise coming from the phase-intensity coupling is also increased for intermediate frequencies, as intuitively anticipated (Fig.\ \ref{Fig05}(a)). But for smaller values of coupling ($ C = 0.1 $ and 0.35), the response of the antiphase relaxation mechanism is lower (for $ C=0.1 $) or comparable ($ C=0.35 $) with the response of the inphase relaxation mechanism for intermediate frequencies (between 200~kHz and 2~MHz). Moreover, the in-phase relaxation mechanism is independent of the strength of coupling. So the noise response of each laser mode, which is obtained by substracting the anti-phase response from the in-phase response, is relatively low at intermediate frequencies for smaller couplings. Therefore, the phase noise introduced via phase-intensity coupling is relatively low at these frequencies in case of low ($ C=0.1 $) and moderate coupling ($ C=0.35 $), which explains the evolution of the experimental and theoretical spectra with $C$. These results suggest that a moderate spatial separation of about 50 $\mu$m, corresponding to roughly  50 \% mode spatial overlap, is a good tradeoff enabling good robustness of the dual frequency oscillation as well as good spectral purity of the beatnote.

\section{Conclusion}
In this study, we study both experimentally and theoretically the phase noise of the RF beatnote generated by optical mixing of two orthogonally polarized modes of the dual-frequency VECSEL. We have investigated the RF phase noise properties for three different coupling strengths between the two laser modes. In the theory, we consider two different effects of pump intensity noise, which we suppose to be the only source of noise in the frequency range of interest (10 kHz to 50 MHz). The first effect of the pump noise is to generate intensity noises in the two laser modes, which then give rise to phase noise via phase-intensity coupling due to the high Henry factor of the semiconductor gain medium. Now the  intensity noises of the two laser modes are not perfectly correlated and the noise correlations depend on the strength of the nonlinear coupling between these modes \cite{De2013}. The partial correlation between the intensity noises of the two laser modes comes from the fact that the pump noises for the two modes are not perfectly correlated ($ \eta < 1$). Therefore, the optical phase noises for the two laser modes, which are generated by intensity noises via phase-intensity coupling, are also not perfectly correlated. As a result, the fluctuations of optical phases of the two laser modes are not completely cancelled out, when we mix them to generate beatnote. This gives rise to the few megahertz wide noise pedestal of the RF beatnote. The other effect of the pump noise is to introduce thermal fluctuations of the semiconductor structure. As a result, the refractive index of the semiconductor and hence the effective cavity length fluctuates, which creates the phase fluctuations of the two modes oscillating inside the cavity. Now again, as the pump fluctuations for the two modes are not fully correlated, the phase fluctuations of the two modes originating via thermal fluctuations are not fully correlated. Therefore, in the beatnote the optical phase noises of the two modes are not completely cancelled out thus giving rise to the RF noise pedestal. In the RF phase noise spectra, the thermal noise  is dominating for lower frequencies (typically below 500 kHz), whereas the noise coming from phase-intensity coupling is dominating for higher offset frequencies (typically within 500 kHz $ - $ 50 MHz). Moreover, the thermal noise is independent of coupling between the two laser modes, whereas the noise coming through phase-intensity coupling depends on coupling strengths. This is due to the fact that intensity noises of the laser modes, which are generating phase noises via phase-intensity coupling, as well as  correlations between these noises depend on coupling strengths between the two laser modes. This study will help us to improve the purity of the optically carried RF signal, which is the very basic requirement for any RADAR application. 

\section*{Acknowledgments}
This work was partially supported by the Agence Nationale de la Recherche (Project NATIF No. ANR-09-NANO-012-01) and by the French RENATECH network. 

\begin{IEEEbiography}[{\includegraphics[width=1in,height=1.25in,clip,keepaspectratio]{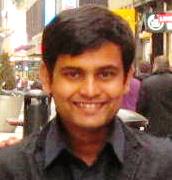}}]{Syamsundar De}
was born in India in 1987. He received his M.S. degree in Applied Physics (Optics, Matter, Plasma) from Ecole Polytechnique, Palaiseau, France, in 2012.

He joined Laboratore Aim\'e Cotton, Orsay, France in 2012 as Ph.D student. He is working on noise of solid-state and semiconductor lasers for microwave photonics applications. His research interests are also in non-linear and quantum optics. 
\end{IEEEbiography}
\begin{IEEEbiography}[{\includegraphics[width=1in,height=1.25in,clip,keepaspectratio]{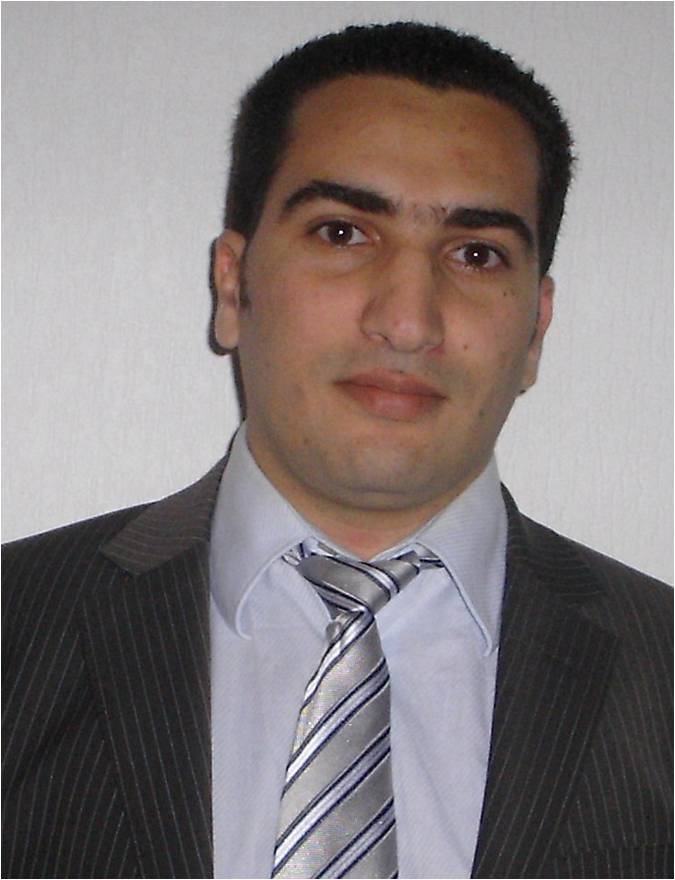}}]{Abdelkrim El Amili}
was born in 1981. He received the M.Sc. from Universit\'e de Bordeaux in 2006 and the PhD degree from Universit\'e Paris Sud 11 in 2009, while working on cold atom manipulation using semiconductor waveguides at Laboratoire Charles Fabry de l'Institut d'Optique. He then joined Laboratoire Aim\'e Cotton to work on VECSELs. In 2011 he joined the Institut de Physique de Rennes where he is working on low noise dual-frequency lasers.
\end{IEEEbiography}
\begin{IEEEbiographynophoto}{Ihsan Fsaifes}
Biography text here.
\end{IEEEbiographynophoto}

\begin{IEEEbiography}[{\includegraphics[width=1in,height=1.25in,clip,keepaspectratio]{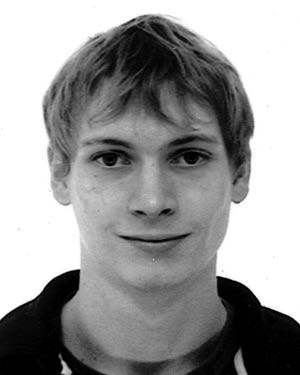}}]{Gr\'egoire Pillet}
received the Master's degree in laser-matter interactions physics from Ecole Polytechnique, France, and the degree from the Ecole Nationale Sup\'erieure des T\'el\'e communication, France, in 2007.
Since 2006, he has been pursuing the Ph.D. degree at the Thales Research and Technology, Palaiseau, France.
\end{IEEEbiography}


\begin{IEEEbiography}[{\includegraphics[width=1in,height=1.25in,clip,keepaspectratio]{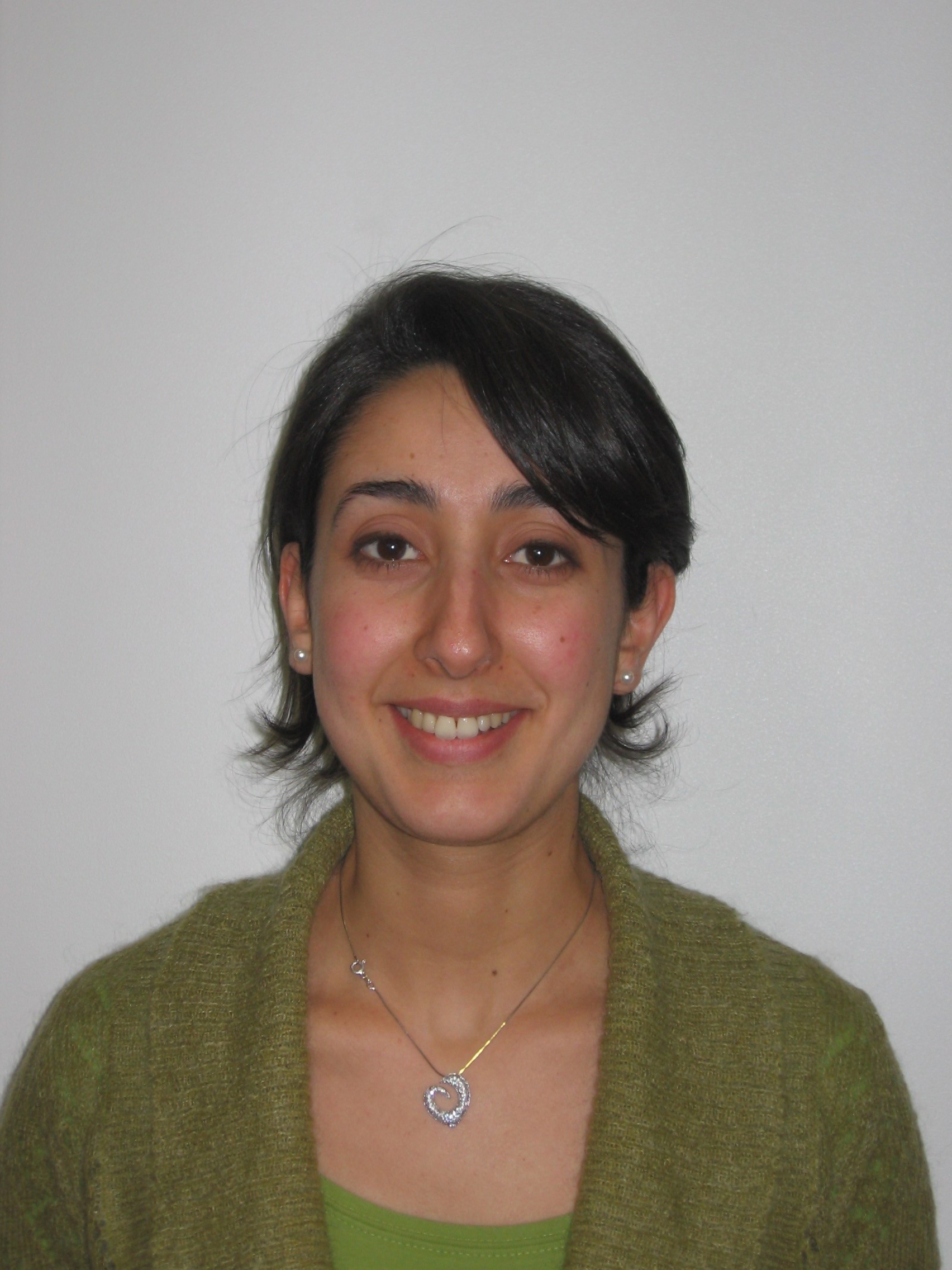}}]{Ghaya Baili}
 was born in Tunisia in 1980. She received the M.S. degree in optics and photonics from the Ecole Sup\'erieure d'Optique, University of Paris XI, Orsay, France, in 2004.
She joined Thales Research and Technology France, Palaiseau, in 2005. She received the PhD degree from Universit\'e Paris Sud 11 in 2008. Her research interests are in the field of lasers for microwave photonics applications.
\end{IEEEbiography}

\begin{IEEEbiography}[{\includegraphics[width=1in,height=1.25in,clip,keepaspectratio]{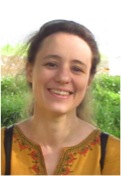}}]{Fabienne Goldfarb}
was born in Paris in 1976. She received the PhD degree from Universit\'e Paris Sud 11 in 2003. She worked on molecular interferometry in Vienna from 2003 to 2005 and then joined Universit\'e Paris Sud 11 as assistant professor in 2005.  Dr. Goldfarb's research interests are in the field of coherent effects in atoms, slow and fast light, and microwave photonics.
\end{IEEEbiography}

\begin{IEEEbiography}[{\includegraphics[width=1in,height=1.25in,clip,keepaspectratio]{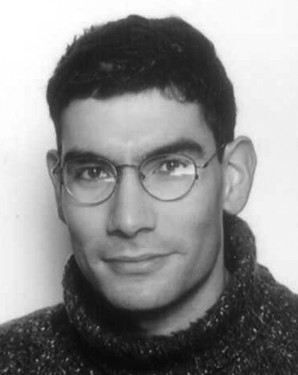}}]{Mehdi Alouini}
received the B.S. degree in fundamental physics from the University of Rennes, Rennes, France, in 1995, the M.S. degree in optics
and photonics from the Ecole Supérieure d'Optique,
University of Paris XI, Orsay, France, in 1997, and
the Ph.D. degree in laser physics from the University
of Rennes in 2001.
He joined Thales Research and Technology
France, Palaiseau, in 2001 as Research Staff
Member. His main activities cover the modeling and
optimization of microwave photonics systems as
well as active polarimetric and multispectral imaging. His research interests are
also in solid-state and semiconductor laser physics for microwave photonics
applications.
\end{IEEEbiography}

\begin{IEEEbiography}[{\includegraphics[width=1in,height=1.25in,clip,keepaspectratio]{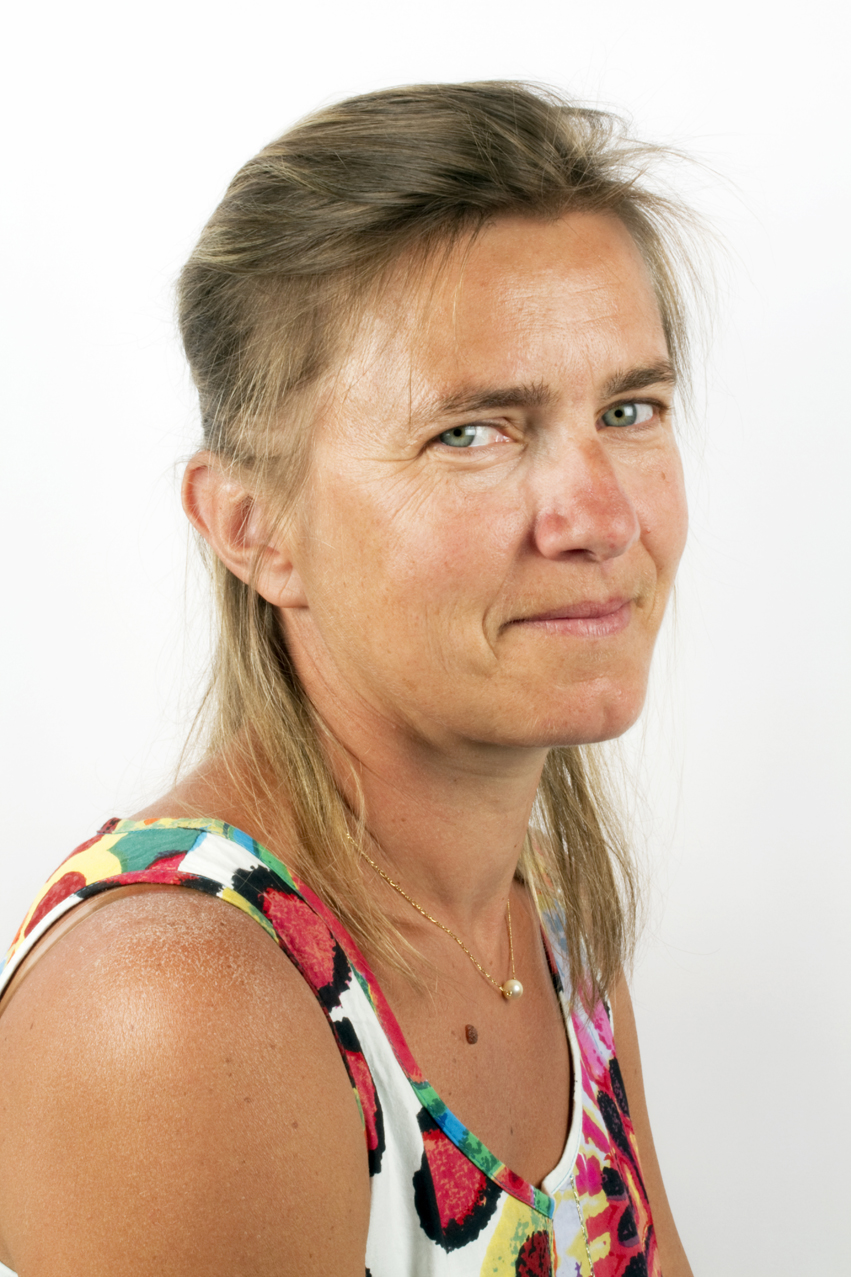}}]{Isabelle Sagnes}
was born in Tunisia in 1967. She received the Ph.D. degree in physics from the University Joseph Fourier, Grenoble, France, in 1994, for her work in electro-optical properties of epitaxial heterostructures on silicon.
From 1994 to 1997, she was with France Telecom/CNET-CNS, Grenoble, as a Process Engineer in silicon microelectronics technology, with a special interest in CVD growth of epitaxial Si/SiGe heterostructures, rapid thermal processing (oxidation, nitridation, implant annealing), and
bipolar and CMOS technologies. She joined the CNET Laboratory, Bagneux, France, in 1998 as a Research Engineer in semiconductor III-V compounds and in 1999 the CNRS in the FT/CNRS joint laboratory (LPN: Laboratoire de Photonique et Nanostructures), Marcoussis, France. The main fields of her research are MOCVD growth of vertical cavity systems on GaAs and InP substrates, MOCVD growth of quantum dots on GaAs and InP substrates, heteroepitaxy of GaAs on Ge/Si virtual substrate, and III-V device technology as the processing of photonic crystals. She is an author or coauthor of about 200 papers in international journals.
\end{IEEEbiography}

\begin{IEEEbiography}[{\includegraphics[width=1in,height=1.25in,clip,keepaspectratio]{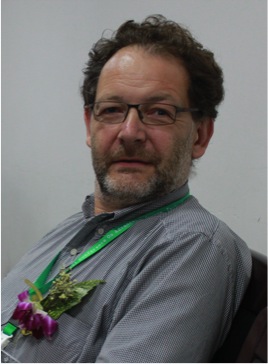}}]{Fabien Bretenaker}
(M'06) was born in Metz, France, in 1966. After having graduated from Ecole Polytechnique, Palaiseau, France, he received the Ph.D. degree from the University of Rennes, Rennes, France, in 1992 while working on ring laser gyroscopes for Sagem.

He joined the Centre National de la Recherche Scientifique, Rennes, in 1994 and worked in Rennes until 2002 on laser physics and nonlinear optics. In 2003, he joined the Laboratoire Aim\'e Cotton, Orsay, France, working on nonlinear optics, laser physics, quantum optics, and microwave photonics. He teaches quantum physics and laser physics in Ecole Polytechnique, Palaiseau, France. He is also an Associate Editor of the European Physical Journal Applied Physics.
Dr. Bretenaker is a member of the Soci\'et\'e Fran\c caise de Physique, the Soci\'et\'e Fran\c caise d'Optique, the Optical Society of America, and IEEE Photonics.
\end{IEEEbiography}




\begin{thebibliography}{1}

\bibitem{Tonda2006}
S.~Tonda-Goldstein, D.~Dolfi, A.~ Monsterl\'e, S.~Formont, J.~Chazelas, and J.-P.~ Huignard, ``Optical signal processing in radar systems," \emph{IEEE Trans. Microw. Theory and Tech.}, vol.~54, no.~6, pp.~847-853, Feb.~2006. 
\bibitem{Rideout2007}
H.~Rideout, J.~Seregelyi, and J.~Yao, ``A true time delay beamforming system incorporating a wavelength tunable optical phase-lock loop," \emph{J. Lightwave Technol.}, vol.~25, no.~27, pp.~1761-1770, Jul.~2007.
\bibitem{Alouini2001}
M.~Alouini, B.~Benazet, M.~Vallet, M.~Brunel, P.~Di~Bin, F.~Bretenaker, A.~Le~Floch, and P.~Thony, ``Offset phase locking of Er:Yb:Glass laser eigenstates for RF photonics applications," \emph{IEEE Photon. Tech. Lett.}, vol.~13, Iss.~4, pp.~367-369, Apr.~2001.
\bibitem{Narbonneau2006}
F.~Narbonneau, M.~Lours, S.~Bize, A.~Clairon, G.~Santarelli, O.~Lopez, C.~Daussy, A.~Amy-Klein, and C.~Chardonnet, ``High resolution frequency standard dissemination via optical fiber metropolitan network," \emph{Rev. Sci. Instrum.}, vol.~77, no.~6, p.~064701,~2007.
\bibitem{Knappe2004}
S.~Knappe, V.~Shah, P.~D.~D.~Schwindt, L.~Hollberg, J.~Kitching, L.-A.~Liew, and J.~Moreland, ``A microfabricated atomic clock," \emph{Appl. Phys. Lett.}, vol.~85, Iss.~9, pp.~1460-1462, Aug.~2007.
\bibitem{Scott1992}
D.~C.~Scott, D.~V.~Plant, and H.~R.~Fetterman, ``60 GHz sources using optically driven heterojunction bipolar transistors," \emph{Appl. Phys. Lett.}, vol.~61, Iss.~1, pp.~1-3, Jul.~1992.
\bibitem{Seeds2006}
A.~J.~Seeds, and K.~J.~Williams, ``Microwave photonics," \emph{J. Lightwave Technol.}, vol.~24, No.~12, pp.~4628-4641, Dec.~2006.
\bibitem{Brunel1997}
M.~Brunel, F.~Bretenaker, and A.~Le Floch, ``Tunable optical microwave source using spatially resolved laser eigenstates," \emph{Opt. Lett.}, vol.~22, No.~6, pp.~384-386, Mar.~1997.
\bibitem{Alouini1998}
M.~Alouini, M.~Brunel, F.~Bretenaker, M.~Vallet, and A.~ Le Floch, ``Dual tunable wavelength Er:Yb:Glass laser for terahertz beat frequency generation," \emph{IEEE Photon. Tech. Lett.}, vol.~10, No.~11, pp.~1554-1556, Nov.~1998.
\bibitem{Czarny2004}
R.~Czarny, M.~Alouini, C.~Larat, M.~Krakowski, and D.~Dolfi, ``THz-dual-frequency Yb$ ^{3+} $:KGd(WO$ _{4} $)$ _{2} $ laser for continuous wave THz generation through photomixing," \emph{Electron. Lett.}, vol.~40, Iss.~15, pp.~942-943, Jul.~2004.
\bibitem{Arecchi1984}
F.~T.~Arecchi, G.~L.~Lippi, G.~P.~Puccioni, and J.~R.~Tredicce, ``Deterministic chaos in laser with injected signal," \emph{Opt. Commun.}, vol.~51, pp.~308-314, 1984.
\bibitem{Taccheo1996}
S.~Taccheo, P.~Laporta, and O.~Svelto, ``Intensity noise reduction in a single-frequency ytterbium-codoped erbium laser," \emph{Opt. Lett.}, vol.~21, No.~21, pp.~1747-1749, Nov.~1996.
\bibitem{Baili2009}
G.~Baili, L.~Morvan, M.~Alouini, D.~Dolfi, F.~Bretenaker, I.~Sagnes and A.~Garnache, ``Experimental demonstration of tunable dual-frequency semiconductor laser free of relaxation oscillation," \emph{Opt. Lett.}, vol.~34, No.~21, pp.~3421-3423, Nov.~2009.
\bibitem{De2013}
S.~De, V.~Pal, A.~El Amili, G.~Pillet, G.~Baili, M.~Alouini, I.~Sagnes, R.~Ghosh, and F.~Bretenaker, ``Intensity noise correlations in a two-frequency VECSEL," \emph{Opt. Expr.}, vol.~21, No.~3, pp.~2538-2550, Jan.~2013.
\bibitem{Henry1982}
C.~H.~Henry, ``Theory of the linewidth of semiconductor lasers," \emph{IEEE J. Quantum Electron.}, vol.~18, No.~2, pp.~259-264, Feb.~1982.
\bibitem{Henry1986}
C.~H.~Henry, ``Phase noise in semiconductor lasers," \emph{J. Lightwave Technol.}, vol.~4, No.~3, pp.~298-311, Mar.~1986.
\bibitem{Agrawala1989}
G.~P.~Agrawal, ``Intensity dependence of the linewidth enhancement factor and its implications for semiconductor lasers," \emph{IEEE Photon. Tech. Lett.}, vol.~1, No.~8, pp.~212-214, Aug.~1989.
\bibitem{Horak2006}
P.~Horak, N.~Y.~Voo, M.~Ibsen, and W.~H.~Loh, ``Pump-noise-induced linewidth contributions in distributed feedback fiber lasers," \emph{IEEE Photon. Tech. Lett.}, vol.~18, no.~9, pp.~998-1000, May.~2006.
\bibitem{Lamb1989}
M.~Sargent III, M.~O.~Scully and W.~E.~Lamb, \emph{Laser Physics}: Addison-Wesley,~1974.
\bibitem{Miguel1995}
M.~San Miguel, Q. Feng, and J.~V.~Moloney, ``Light-polarization dynamics in surface-emitting semiconductor lasers," \emph{Phys. Rev. A}, vol.~52, No.~2, pp.~1728-1739, Aug.~1995.
\bibitem{Travagnin1997}
M.~Travagnin, M.~P.~van Exter, and J.~P.~Woerdman, ``Influence of carrier dynamics on the polarization stability and noise-induced polarization hopping in surface-emitting semiconductor lasers," \emph{Phys. Rev. A}, vol.~56, no.~2, pp.~1497-1507, Aug.~1997.
\bibitem{Exter1998}
M.~P.~van Exter, A.~Al-Remawi and J.~P.~Woerdman, ``Polarization fluctuations demonstrate nonlinear anisotropy of a vertical-cavity semiconductor laser," \emph{Phys. Rev. Lett.}, vol.~80, no.~22, pp.~4875-4878, Jun.~1998.
\bibitem{Gahl1999}
A.~Gahl, S.~Balle, and M.~San Miguel, ``Polarization dynamics of optically pumped VCSEL's," \emph{IEEE J. Quantum Electron.}, vol.~35, no.~3, pp.~342-351, Mar.~1999.
\bibitem{Kaiser2002}
J.~Kaiser, C.~Degen, and W.~Elsasser, ``Polarization-switching influence on the intensity noise of vertical-cavity surface-emitting lasers," \emph{J. Opt. Soc. Am. B}, vol.~19, no.~4, pp.~672-677, Apr.~2002.
\bibitem{Tropper2006}
A. C. Tropper and S. Hoogland, ``Extended cavity surface-emitting semiconductor lasers," \emph{Progr. Quantum Electron.}, vol.~30, pp.~1-43, 2006.
\bibitem{Laurain2009}
A.~Laurain, M.~Myara, G.~Beaudoin, I.~Sagnes and A.~Garnache, ``High power single-frequency continuously-tunable compact extended-cavity semiconductor laser," \emph{Opt. Expr.}, vol.~17, no.~12, pp.~9503-9508, Jun.~2009.
\bibitem{Laurain2010}
A.~Laurain, M.~Myara, G.~Beaudoin, I.~Sagnes and A.~Garnache, ``Multiwatt-power highly-coherent
compact single-frequency tunable Vertical-External-Cavity-Surface-Emitting-Semiconductor-Laser," \emph{Opt. Expr.}, vol.~18, no.~14, pp.~14627-14636, Jul.~2010.
\bibitem{LaurainThese}
A. Laurain, ``Sources laser à semiconducteur à \'emission verticale de haute coh\'erence et de forte puissance dans le proche et le moyen infrarouge,'' \emph{PhD thesis}, Univ. Montpellier II, 2010.
\bibitem{Foster2004}
S. Foster, ``Dynamical noise in single-mode distributed feedback fiber lasers," \emph{IEEE J. Quantum Electron.}, vol.~40, no.~9, pp.~1283-1293, Sep.~2004.
\bibitem{Foster2008}
S. Foster, ``Fundamental limits on 1/f noise in rare-earth-metal-dopped fiber lasers due to spontaneous emission," \emph{Phys. Rev. A}, vol.~78, pp.~013820-1  - 013820-7, Jul.~2008.
\bibitem{Fukuda1993}
M.~Fukuda, T.~Hirono, T.~Kurosaki, and F.~Kano, ``1/f noise behavior in semiconductor laser degradation," \emph{IEEE Photon. Tech. Lett.}, vol.~5, no.~10, pp.~1165-1167, Oct.~1993.
\bibitem{Davis1998}
M.~K.~Davis, M.~J.~F.~Digonnet, and R.~H.~Pantel, ``Thermal effects in dopped fibers," \emph{J. Lightwave Tech.}, vol.~16, no.~6, pp.~1013-1023, Jun.~1998.
\bibitem{Chui1992}
T.~C.~P.~Chui, D.~R.~Swanson, M.~J.~Adriaans, J.~A.~Nissen, and J.A~.~Lipa, ``Temperature fluctuations in the canonical ensemble," \emph{Phys. Rev. Lett.}, vol.~69, no.~21, pp.~3005-3008, Nov.~1992.
\bibitem{Gorodetsky2004}
M.~L.~Gorodetsky, and I.~S.~Grudinin, ``Fundamental thermal fluctuations in microspheres," \emph{J. Opt. Soc. Am. B}, vol.~21, no.~4, pp.~697-705, Apr.~2004.
\bibitem{Lauer2005}
C.~Lauer, and M.~-C.~Amann, ``Calculation of the linewidth broadening in vertical-cavity surface-emitting lasers due to temperature fluctuations," \emph{Appl. Phys. Lett.}, vol.~86, pp.~191108-1 - 191108-3, 2005.

\bibitem{Pal2010}
V.~Pal, P.~Trofimoff, B.~-X.~Miranda, G.~Baili, M.~Alouini, L.~Morvan, D.~Dolfi, F.~Goldfarb, I.~Sagnes, R.~Ghosh and F.~Bretenaker, ``Measurement of the coupling constant in a two-frequency VECSEL," \emph{Opt. Expr.}, vol.~18, no.~5, pp.~5008-5014, Mar.~2010.
\end{thebibliography}
\end{document}